\documentclass[
preprint,
showpacs,preprintnumbers,
bibnotes,
 amsmath,amssymb,
 aps,
 pra,
superscriptaddress,
longbibliography,
]{revtex4-1}

\usepackage[version=3]{mhchem} 
\usepackage[dvipsnames]{xcolor}
\usepackage{textcomp}
\usepackage{graphicx}
\usepackage{amsmath}
\usepackage{fixmath}
\usepackage{amsfonts}
\usepackage{amssymb}

\newcommand{\micron}{\hbox{\textmu}\text{m}}

\graphicspath{ {./figures/} }

\begin{document}

\title{
Nonlinear polaritons in monolayer semiconductor coupled to optical bound states in the continuum
}

\author{Vasily Kravtsov}
\email{Corresponding author: vasily.kravtsov@metalab.ifmo.ru}
\altaffiliation{Contributed equally}
\author{Ekaterina Khestanova}
\altaffiliation{Contributed equally}
\author{Fedor A. Benimetskiy}
\altaffiliation{Contributed equally}
\author{Tatiana Ivanova}
\author{Anton K. Samusev}
\author{Ivan S. Sinev}
\author{Dmitry Pidgayko}
\affiliation{ITMO University, Saint Petersburg 197101, Russia}
\author{Alexey M. Mozharov}
\affiliation
{St. Petersburg Academic University, Saint Petersburg 194021, Russia}
\author{Ivan S. Mukhin}
\affiliation{ITMO University, Saint Petersburg 197101, Russia}
\affiliation
{St. Petersburg Academic University, Saint Petersburg 194021, Russia}
\author{Maksim S. Lozhkin}
\author{Yuri V. Kapitonov}
\affiliation
{Saint Petersburg State University, ul. Ulyanovskaya 1, Saint Petersburg 198504, Russia}
\author{Andrey S. Brichkin}
\affiliation
{Institute of Solid State Physics, RAS, Chernogolovka 142432, Russia}
\author{Vladimir D. Kulakovskii}
\affiliation
{Institute of Solid State Physics, RAS, Chernogolovka 142432, Russia}
\author{Ivan A. Shelykh}
\affiliation{ITMO University, Saint Petersburg 197101, Russia}
\affiliation{Science Institute, University of Iceland, Dunhagi 3, IS-107, Reykjavik, Iceland}
\author{Alexander I. Tartakovskii}
\author{Paul M. Walker}
\affiliation
{Department of Physics and Astronomy, University of Sheffield, Sheffield S3 7RH, UK}
\author{Maurice S. Skolnick}
\affiliation{ITMO University, Saint Petersburg 197101, Russia}
\affiliation
{Department of Physics and Astronomy, University of Sheffield, Sheffield S3 7RH, UK}
\author{Dmitry N. Krizhanovskii}
\affiliation
{Department of Physics and Astronomy, University of Sheffield, Sheffield S3 7RH, UK}
\author{Ivan V. Iorsh}
\email{Corresponding author: i.iorsh@metalab.ifmo.ru}
\affiliation{ITMO University, Saint Petersburg 197101, Russia}

\date{\today}



\begin{abstract}
\noindent 
\textbf{
Optical bound states in the continuum (BICs) provide a way to engineer very narrow resonances in photonic crystals~\cite{Marinica2008,Hsu2016,Lee2012,Jin2018}.
The extended interaction time in such systems is particularly promising for enhancement of nonlinear optical processes~\cite{Krasikov2018,Bulgakov2019} and development of the next generation of active optical devices~\cite{Walker2015,Sich2016}.
However, the achievable interaction strength is limited by the purely photonic character of optical BICs.
Here, we mix optical BIC in a photonic crystal slab with excitons in atomically thin semiconductor MoSe$_2$ to form nonlinear exciton-polaritons with a Rabi splitting of 27~meV, exhibiting large interaction-induced spectral blueshifts.
The asymptotic BIC-like suppression of polariton radiation into far-field towards the BIC wavevector, in combination with effective reduction of excitonic disorder through motional narrowing, results in small polariton linewidths below 3~meV.
Together with strongly wavevector-dependent Q-factor, this provides for enhancement and control of polariton--polariton interactions and resulting nonlinear optical effects, paving the way towards tunable BIC-based polaritonic devices for sensing, lasing, and nonlinear optics.
}
\end{abstract}

\maketitle

\noindent Optical bound states in the continuum (BICs), supported by photonic crystal structures of certain geometries, have received much attention recently as a novel approach to generating extremely spectrally narrow resonant responses~\cite{Marinica2008,Hsu2016}.
Since BICs are uncoupled from the radiation continuum through symmetry protection~\cite{Lee2012} or resonance trapping~\cite{Kodigala2017}, they can be robust to perturbations of photonic crystal geometric parameters~\cite{Jin2018}.
This enables a broad range of practical applications, including recently demonstrated spectral filtering~\cite{Foley2014}, chemical and biological sensing~\cite{Romano2018a,Romano2018b}, and lasing~\cite{Kodigala2017}.

Providing an efficient light-trapping mechanism, optical BICs are particularly attractive for enhancing nonlinear optical effects, with recent theoretical proposals discussing enhanced bistability~\cite{Bulgakov2019} and Kerr-type focusing nonlinearity~\cite{Krasikov2018}.
However, for practical realization of these proposals, a significantly stronger material nonlinear susceptibility is needed than generally available in dielectric-based photonic crystals.

An attractive approach to the enhancement of effective nonlinearity is through the use of exciton-polaritons -- hybrid quasi-particles that inherit both the coherent properties of photonic modes and interaction strength of excitons~\cite{Khitrova1999,Walker2015,Sich2016}.
Hybrid nanophotonic systems incorporating atomically thin transition metal dichalcogenides (TMDs) have emerged as a particularly promising platform owing to their ease of fabrication and possibility of room temperature operation~\cite{Dufferwiel2015,Liu2015,Lundt2016}. In addition to conventional microcavity-based designs, TMD exciton-polaritons have been observed in plasmonic lattices~\cite{Dibos2019}, photonic crystal slabs~\cite{Zhang2018,Gogna2019}, and other nanophotonic structures~\cite{Barachati2018}. 

Coupling TMD excitons to optical BICs in photonic crystals thus will not only boost the potentially achievable nonlinearities but also provide control on the resonant BIC properties through the excitonic fraction in the polariton, as has been proposed theoretically~\cite{Koshelev2018}.

Here we experimentally demonstrate and investigate nonlinear polaritons formed via strong coupling of excitons in monolayer MoSe$_2$ and optical BIC in a 1D photonic crystal slab, with Rabi splitting of $>27$~meV and BIC-like radiation suppression in the surface-normal direction.
Despite the large $\sim 9$~meV inhomogeneous broadening of the MoSe$_2$ excitonic line, we achieve small polariton linewidth below 3~meV, corresponding to very well resolved splitting-to-linewidth ratio of $\sim 9$ and Q-factors up to 900.
Using the strongly wavevector-dependent Q-factor of the photonic crystal dispersion, we show controllable reduction in polariton linewidth by a factor of 5-10 when approaching the BIC.
The narrow polariton lines allow us to accurately measure polariton--polariton interaction strength through power-dependent spectral blueshifts in resonant reflectance experiment, corresponding to exciton--exciton interaction strength of $g_\mathrm{X} \sim 1.0$~\textmu{}eV$\cdot$\textmu{}m$^2$.
This polariton nonlinearity is comparable to the values observed in III-V materials~\cite{Ferrier2011,Sich2012} and significantly larger than previously observed in atomically thin semiconductors~\cite{Barachati2018}, paving the way towards quantum applications of exciton-polaritons in atomically thin semiconductors.


\begin{figure*}[tb]
	\includegraphics[width=\columnwidth]{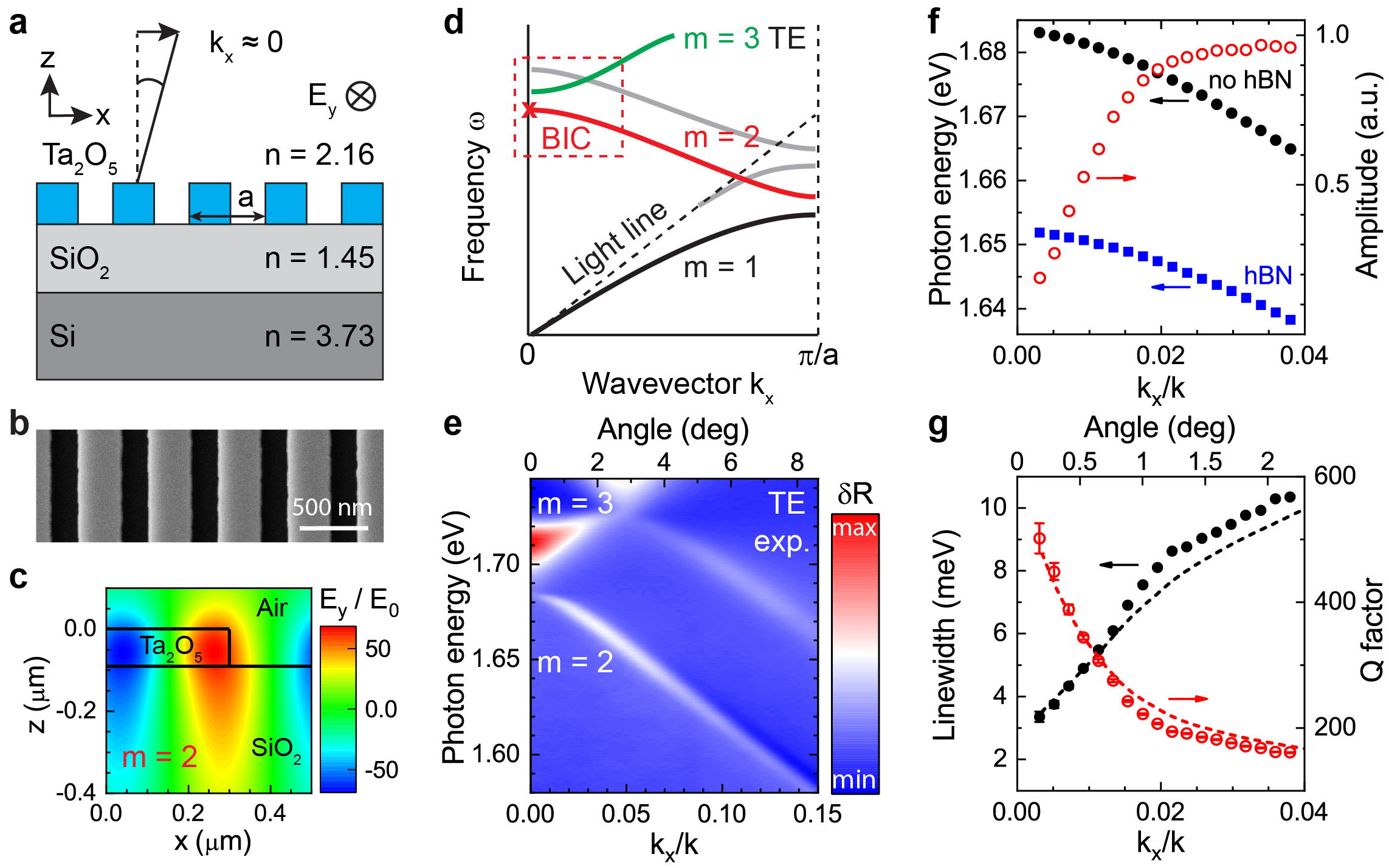}
	\caption{\textbf{At-$\Gamma$ optical BIC in a photonic crystal slab.} (a) Schematic of photonic crystal slab (PCS) sample, with Ta$_2$O$_5$ bars on SiO$_2$/Si substrate, illuminated with TE-polarized light near normal incidence. (b) SEM image of PCS sample. (c) Calculated TE mode electric field distribution at $k_x/k = 0.024$. (d) Schematic of photonic band structure for TE modes, with at-$\Gamma$ optical BIC position indicated with X. (e) Experimental differential angle-resolved reflectance spectra showing one symmetric and two antisymmetric modes. (f) Wavevector-dependent peak position (black) and amplitude (red) extracted using Fano-like fits for antisymmetric mode near BIC location; blue squares show peak positions shifted due to a 9~nm hBN layer. (g) Extracted wavevector-dependent linewidth (black) and Q-factor (red), together with corresponding simulation results (dashed lines) corrected for experimental resolution and scattering losses.}
	\label{fig:BIC}
\end{figure*}

In the experiment, we fabricate a 1D photonic crystal slab (PCS) sample consisting of 90~nm thick Ta$_2$O$_5$ bars on a SiO$_2$/Si (1~\micron /500~\micron) substrate as schematically shown in Fig.~\ref{fig:BIC}a, with a scanning electron microscopy image in (b).
The PCS geometry (see Methods) is designed for large refractive index modulation, in order to open a photonic band gap and support an optical BIC close to the exciton resonance in monolayer MoSe$_2$.
As illustrated in the photonic band structure shown in Fig.~\ref{fig:BIC}d, the BIC is expected to form on the lower-energy $m=2$ TE mode (red) at the $\Gamma$ point in the crystal momentum ($k$) space~\cite{Koshelev2018}, with characteristic confinement and antisymmetric spatial distribution of the optical field (c).

We measure the PCS band structure via angle-resolved reflectance spectroscopy (see Methods and Supplementary Fig.~S1).
Fig.~\ref{fig:BIC}e shows experimental differential reflectance spectra for varying angle, where the signal from the un-patterned Ta$_2$O$_5$/SiO$_2$/Si substrate is subtracted for clarity: $\delta R(\alpha, \omega) = R_\mathrm{PCS}(\alpha, \omega) - R_\mathrm{sub}(\alpha, \omega)$.
Three modes, one broad symmetric ($m=3$) and two narrower antisymmetric ($m=2$), are clearly observed in the figure, in agreement with theory (d, red dashed box).
We fit the lower-energy antisymmetric mode peak in reflectance spectra using a Fano-like line shape $F(\omega) \propto (q\gamma / 2 + \omega - \omega_0)^2/(\gamma^2 / 4 + (\omega - \omega_0)^2)$ with resonance frequency $\omega_0$, linewidth $\gamma$, and asymmetry parameter $q$, which arises due to interference with the broad symmetric mode and even broader Fabry-Perot response of the layered substrate (see Supplementary Fig.~S2).

Extracted Fano fit parameters are plotted in Fig.~\ref{fig:BIC}f,g as functions of the in-plane wavevector component $k_x/k$.
Towards the $\Gamma$ point ($k_x \to 0$), the reflectivity associated with the mode sharply decreases (f, red), while the spectral line narrows (g, black circles) resulting in sharply increasing Q-factor (g, red open circles), defined as $Q = \omega_0/\gamma$.
Such behavior is characteristic of an at-$\Gamma$ optical BIC~\cite{Hsu2013}, where interference of optical waves outgoing in opposite directions leads to effective light trapping in the near field and vanishing far-field radiation.
This is in contrast to the case studied recently~\cite{Zhang2018,Gogna2019}, where only radiating PCS modes with smaller and largely angle-independent Q-factors were considered for strong coupling to excitons in 2D semiconductors. 

\begin{figure*}[tb]
	\includegraphics[width=\columnwidth]{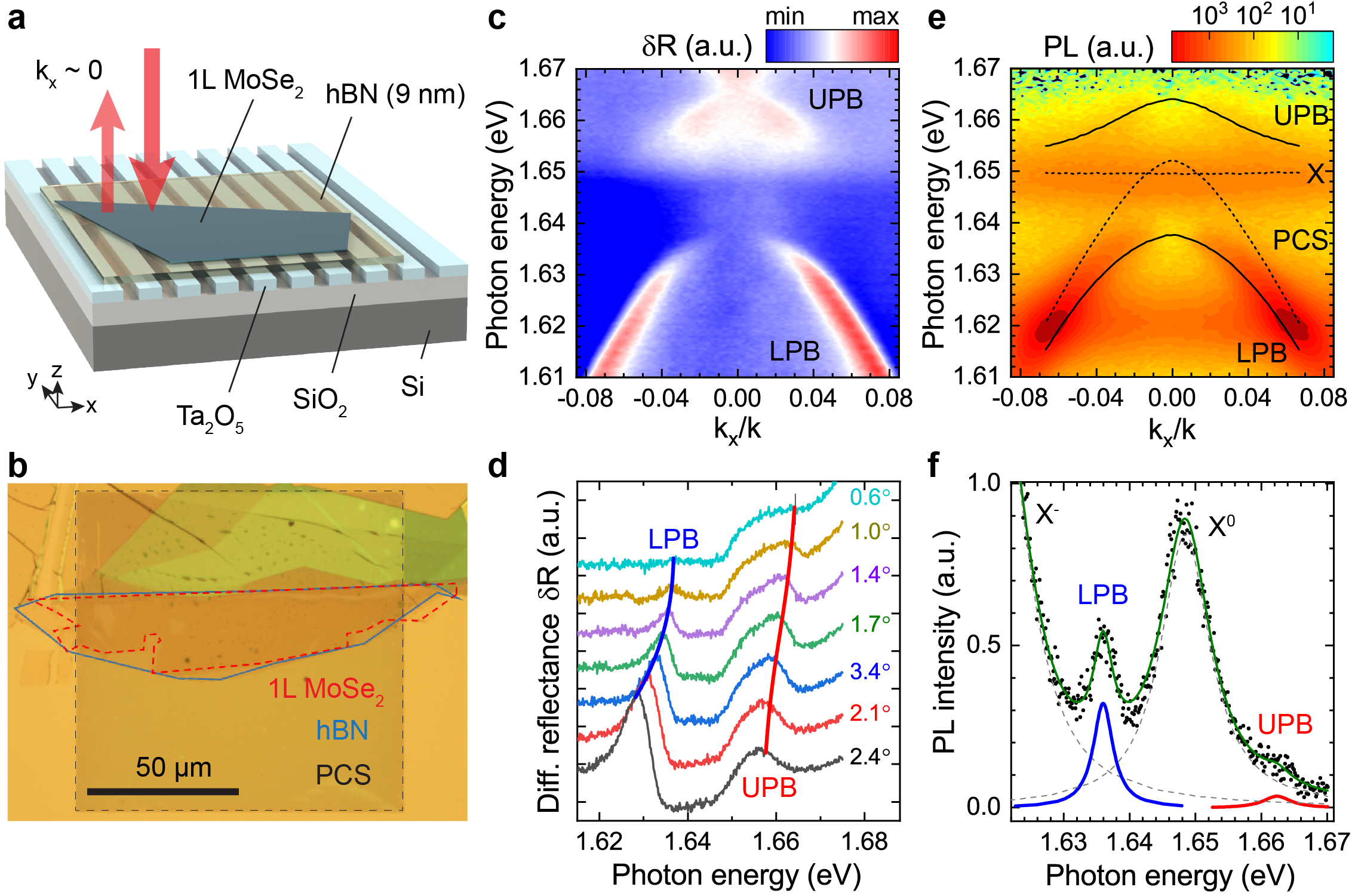}
	\caption{\textbf{Strong coupling of excitons in 1L MoSe$_2$ and BIC.} (a) Schematic of hybrid 1L MoSe$_2$/hBN/PCS structure. (b) Optical microscope image of the fabricated sample. (c) Angle-resolved reflectance spectra of the hybrid sample showing upper and lower polariton branches due to strong coupling between MoSe$_2$ exciton and antisymmetric TE mode of PCS. (d) Differential reflectance spectra for selected angles. (e) Angle-resolved PL spectra (TE-polarized) with indicated positions of uncoupled PCS mode and MoSe$_2$ exciton (dashed lines) and resulting polaritons (solid lines). (f) Experimental PL spectrum for $k_x/k = 0.015$ (dots) and Lorentz fits. Dashed lines indicate fits for uncoupled exciton and trion peaks, blue and red lines indicate fits for lower and upper polariton peaks, respectively.}
	\label{fig:Dispersion}
\end{figure*}

The theoretically predicted \textit{radiative} Q-factor of BIC diverges towards infinity at $k_x=0$.
In practice, the measured Q-factor is limited predominantly by \textit{non-radiative} losses~\cite{Hsu2013,Jin2018}.
The two major contributions in our case are (i) intrinsic absorption in Ta$_2$O$_5$ and (ii) surface roughness induced symmetry breaking and scattering~\cite{Sadrieva2017}, limiting Q to $\sim 10^3$.
Additional resonance broadening comes from leaky losses in Si due to near-field penetration through SiO$_2$ layer~\cite{Sadrieva2017} and finite sample size effects~\cite{Grepstad2013, Lee2012}.
We perform Fourier modal method (FMM) simulations, taking into account these four loss mechanisms (see Supplementary Note~1 and Fig.~S3). 
As seen in Fig.~\ref{fig:BIC}g (dashed lines), good agreement with experiment can be achieved considering scattering losses through an additional imaginary part of Ta$_2$O$_5$ refractive index $\delta n \sim 0.002i$ (Fig.~\ref{fig:BIC}g, dashed lines).

We then create polaritons by coupling the observed BIC to excitons in a vertically stacked 1L MoSe$_2$/hBN/PCS structure illustrated in Fig.~\ref{fig:Dispersion}a.
To maximize the Q-factor of resulting polariton modes, we use large-area multilayer hBN and monolayer MoSe$_2$ flakes of $\sim 100$~\micron~in size, covering $\sim 200$ periods of PCS, as demonstrated in an optical microscope image in (b).
The hBN spacer plays a three-fold role: it avoids MoSe$_2$ flake ``sagging" into PCS grooves, reduces influence of Ta$_2$O$_5$ surface roughness, and provides tunability of the BIC frequency through hBN thickness.
In our case, a 9~nm thick hBN spacer shifts the PCS mode and spectral position of BIC by $\sim 30$~meV to bring it close to resonance with the neutral exciton in 1L MoSe$_2$ at 7~K $\hbar\omega_\mathrm{X} = 1.65$~eV (Fig.~\ref{fig:BIC}f, blue squares).

We study polaritons experimentally via angle-resolved reflectivity and PL measurements at 7~K (see Methods), with the results for TE-polarized detection shown in Fig.~\ref{fig:Dispersion}c-f.
In comparison to Fig.~\ref{fig:BIC}e, the lower-energy antisymmetric PCS mode observed in reflectivity (Fig.~\ref{fig:Dispersion}c,d) is now redshifted by $\sim 30$~meV due to the presence of hBN/MoSe$_2$ and split into upper and lower polariton branches (UPB, LPB) due to strong coupling with the neutral exciton in 1L MoSe$_2$ centered at $\hbar\omega_\mathrm{X} = 1.65$~eV.
Both LPB and UPB retain BIC-like behavior near the $\Gamma$ point, exhibiting several distinctive properties.

\begin{figure*}[tb]
	\includegraphics[width=\columnwidth]{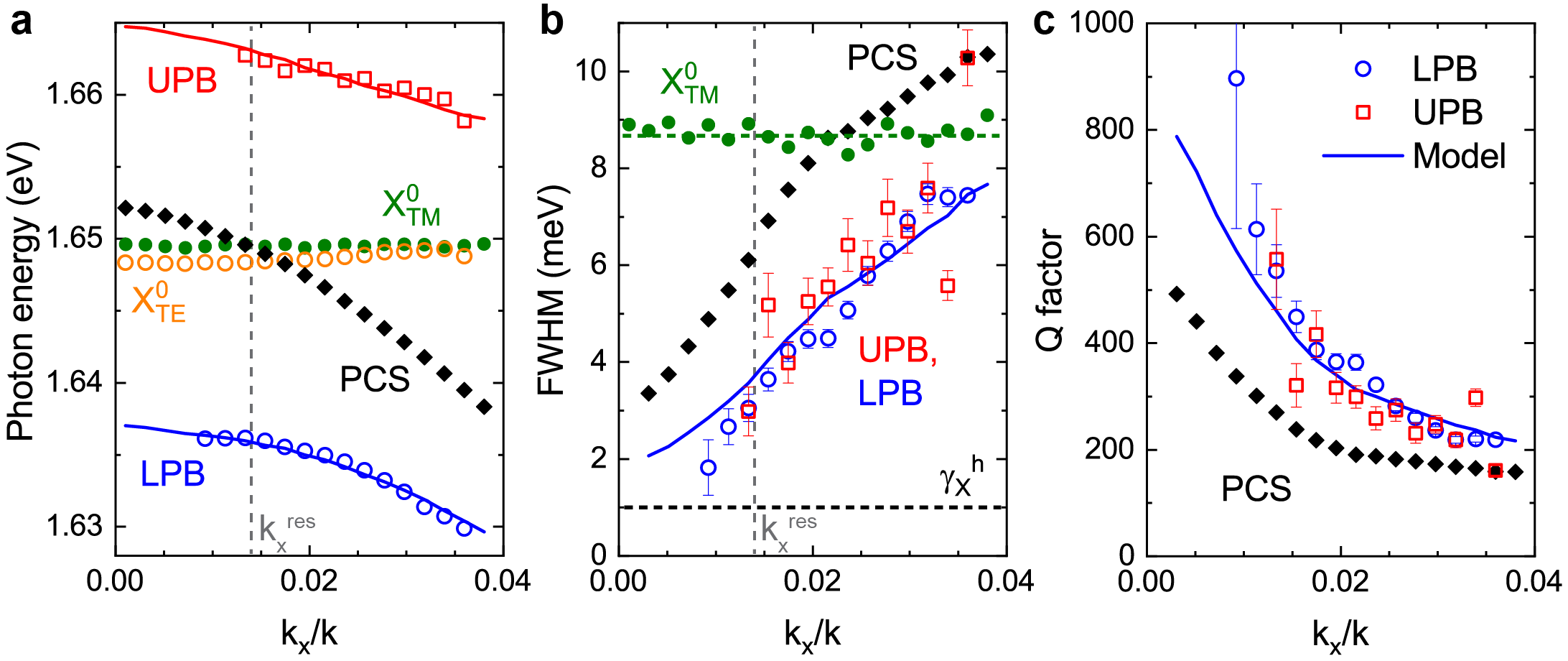}
	\caption{\textbf{Motional narrowing for BIC-based exciton-polaritons.} (a) Spectral peak position, (b) linewidth as full width at half maximum (FWHM), and (c) corresponding Q-factor for PCS mode (black diamonds), lower polariton branch (LPB, blue circles), and upper polariton branch (UPB, red squares), extracted from fits and compared to the coupled harmonic oscillator model. Also shown are the parameters of MoSe$_2$ neutral exciton in TM polarization (a-b, green circles), peak position of uncoupled exciton in TE polarization (a, orange circles), and estimated homogeneous linewidth of excitons $\gamma_\mathrm{X}^{h}$ (b, black dashed line). Blue lines in b-c show model calculations for UPB FWHM and Q-factor.}
	\label{fig:Results}
\end{figure*}

First, at the $\Gamma$ point both LPB and UPB are ``dark" as radiation into the far-field becomes symmetry-forbidden, effectively extending interaction time for potential enhancement of optical nonlinearities.
Second, close to the $\Gamma$ point polaritons possess negative effective mass and associated negative group velocity inherited from the PCS mode, providing a potential platform for studying TMD-based polariton self-focusing and soliton formation.
Third, the strong variation of the PCS mode linewidth in the vicinity of the BIC results in wavevector-dependent Q-factor of both LPB and UPB, enabling control of polariton linewidth with angle.

Further details of the optical response are revealed by TE-polarized angle-resolved PL spectra (e), showing emission from both polariton branches and the uncoupled neutral exciton (X$^0$).
The latter is increasingly enhanced towards small wavevectors and exhibits a slight redshift of $\sim 1$~meV, which we attribute to weak coupling to the higher-frequency and broader $m=3$ symmetric mode (see Supplementary Note~2 and Fig.~S4).
Charged exciton (trion, X$^-$) emission is also observed at $\hbar\omega_\mathrm{T} = 1.62$~meV independent of $k_x$, implying weak coupling.

We analyze the wavevector-dependent behavior of LPB, UPB, uncoupled neutral exciton, and trion by fitting the PL spectra for each $k_x$ with 4 Lorentzian functions $L_i(\omega) \propto ((\omega - \omega_i)^2 + (\gamma_i / 2)^2)^{-1}$, as shown Fig.~\ref{fig:Dispersion}f, and extracting the spectral position $\omega_i$ and linewidth $\gamma_i$ as full width at half maximum (FWHM) for each peak.
The extracted parameters are plotted as functions of in-plane wavevector in Fig.~\ref{fig:Results}, with spectral positions (a) for UPB (red symbols), LPB (blue symbols), and uncoupled neutral exciton (orange symbols), corresponding values of FWHM (b), and calculated Q-factors (c).
Parameters of uncoupled excitons, extracted from TM-polarized PL, are plotted as green dots.

We then fit the extracted spectral positions of UPB ($\omega_+$) and LPB ($\omega_-$) with a coupled oscillator model~\cite{Hopfield1958}, using spectral position and homogeneous linewidth for the uncoupled neutral exciton $\tilde\omega_\mathrm{X} = \omega_\mathrm{X} + i \gamma_\mathrm{X}$ and for the PCS/hBN mode $\tilde\omega_\mathrm{C}(k_x) = \delta\omega_\mathrm{C} + \omega_\mathrm{C}(k_x) + i \gamma_\mathrm{C}(k_x)$:
$$
\omega_{\pm} = \frac{\tilde\omega_\mathrm{C}(k_x) + \tilde\omega_\mathrm{X}}{2} \pm \frac{1}{2}\sqrt{\hbar^2 \Omega_\mathrm{R}^2 + (\tilde\omega_\mathrm{C}(k_x) - \tilde\omega_\mathrm{X})^2},
$$
$$
\hbar\Omega_\mathrm{R} = 2\sqrt{g^2 - \frac{(\gamma_\mathrm{C}(k_x) - \gamma_\mathrm{X})^2}{4}}.
$$
Here, $\Omega_\mathrm{R}$ is Rabi splitting between UPB and LPB, and the two fit parameters are the coupling strength $g$ and additional spectral shift $\delta\omega_\mathrm{C}$ of the PCS mode~\cite{Zhang2018} due to the presence of 1L MoSe$_2$.
The fit curves for the spectral position of UPB and LPB are shown in Fig.~\ref{fig:Results}a by red and blue solid lines, respectively.
The uncoupled PCS/hBN mode, indicated by black squares, comes into resonance with the uncoupled neutral exciton at $k_x^\mathrm{res}/k \simeq \pm 0.014$, corresponding to an angle of $\alpha \simeq \pm 0.8^\circ$.
From the fits in Fig.~\ref{fig:Results}a we extract a coupling strength of $g = 13.9$~meV, which corresponds to a Rabi splitting of $\Omega_\mathrm{R} = 27.4$~meV and splitting-to-linewidth ratio of $\sim 9$, exceeding the values recently reported for a WSe$_2$/PCS system~\cite{Zhang2018} and theoretical estimates for strong coupling to an optical BIC~\cite{Koshelev2018}.
As $\Omega_\mathrm{R}$ is larger than the sum of the exciton ($\sim 9$~meV) and PCS mode linewidth ($\sim 3-11$~meV), the hybrid MoSe$_2$/hBN/PCS system is unambiguously in the strong coupling regime.

Quantitatively, the polariton linewidth $\gamma_{\pm}$ is expected to vary between that of the exciton ($\gamma_\mathrm{X}$) and PCS modes ($\gamma_\mathrm{C}$) depending on the excitonic fraction in the polariton.
However, our experimentally observed values of polariton linewidth (Fig.~\ref{fig:Results}b, open circles) close to resonance $k_x = k_x^\mathrm{res}$ are significantly smaller than both $\gamma_\mathrm{X}$ (green) and $\gamma_\mathrm{C}$ (black).
We attribute this to polariton motional narrowing, similar to the effects studied previously for quantum wells in microcavities~\cite{Whittaker1996, Savona1997, Kavokin1998, Whittaker1998, Skolnick1998}.

Here, the large polariton mode size (10's of \micron) together with large Rabi splitting lead to effective averaging over excitonic disorder~\cite{Rhodes2019} in 1L MoSe$_2$ on a broad (nm--\micron) range of length scales.
As a result, the excitonic contribution to the polariton FWHM close to resonance is given by only the \textit{homogeneous} exciton linewidth $\gamma_\mathrm{X}^\mathrm{h}$~\cite{Houdre1996}, while away from resonance, where the polariton frequency overlaps with the exciton peak, it changes towards the \textit{inhomogeneous} linewidth $\gamma_\mathrm{X}^\mathrm{inh}$ due to increasing interaction with disorder and associated scattering with higher-momenta excitonic states as well as absorption~\cite{Whittaker1998, Walker2017}.
We use a phenomenological model that accounts for homogeneous and inhomogeneous contributions to the polariton linewidth (see Supplementary Note~3).
The model shows good agreement with experimental data (Fig.~\ref{fig:Results}b, blue line) for a homogeneous linewidth of $\gamma_\mathrm{X}^\mathrm{h} \sim 1$~meV (b, black dashed line), which is within the range of recently reported values~\cite{Ajayi2017, Scuri2018, Martin2018, Fang2019} for the low-temperature radiative decay rate of excitons in monolayer MoSe$_2$.

As a result of excitonic and photonic disorder averaging, the Q-factors achieved for polaritons around the $\Gamma$ point in our structure are $\sim 2$ times higher than those for the bare PCS mode (Fig.~\ref{fig:Results}c), reaching $Q \sim 900$.
This provides potential for further improvement of polariton linewidth through fabrication of macroscopic photonic crystal samples~\cite{Lee2012} with improved surface quality and use of large TMD flakes grown by chemical vapor deposition.
In addition, the strongly $k$-dependent Q-factor of the PCS mode in the vicinity of the BIC enables precise control of polariton linewidth and corresponding Q-factors via angle or temperature tuning (see Supplementary Note~4 and Figs.~S5,S6).

\begin{figure*}[tb]
	\includegraphics[width=0.95\columnwidth]{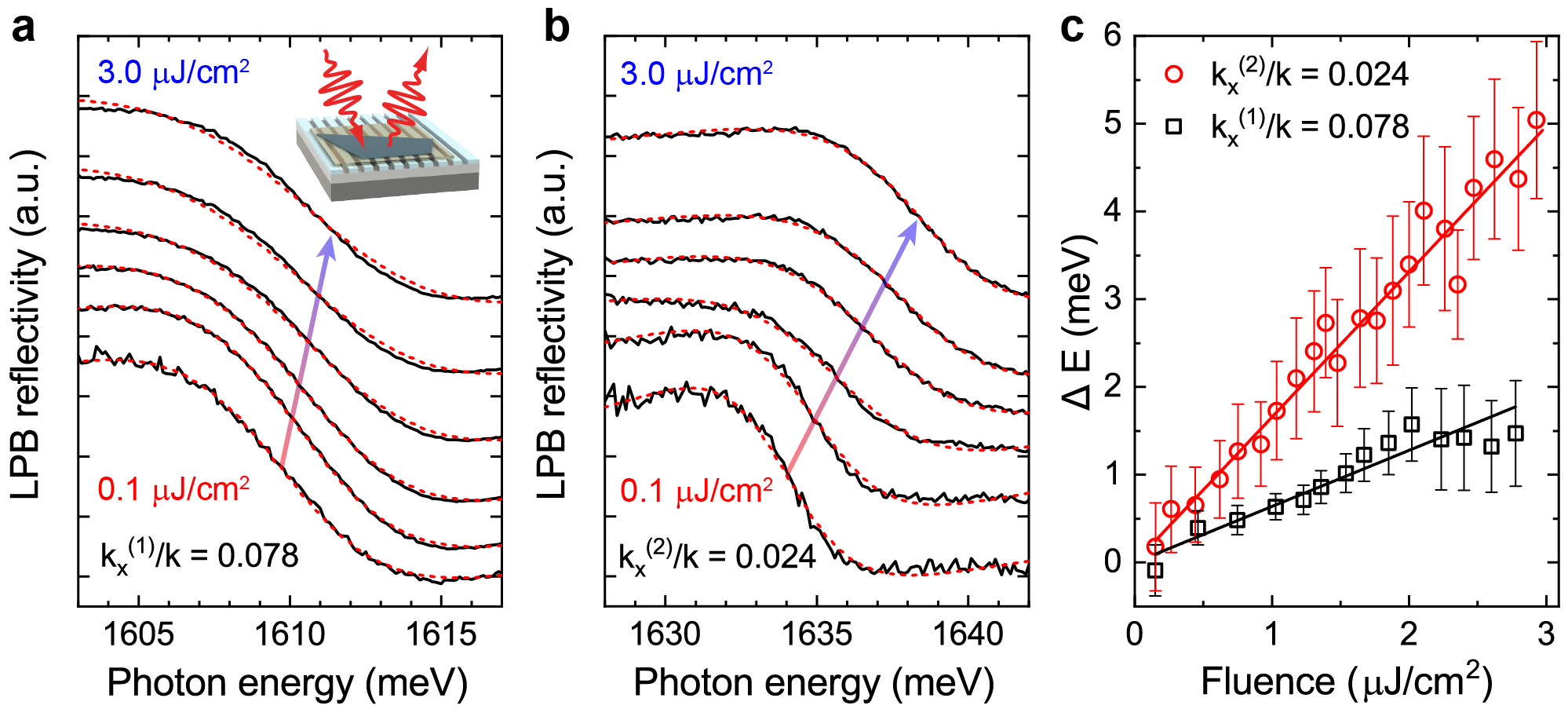}
	\caption{\textbf{Nonlinear interaction of BIC-based polaritons.} (a) Measured LPB reflectance spectra (solid black curves) under resonant illumination with laser pulses at $k_x^{(1)}/k = 0.078$ for varying fluence, together with corresponding Fano fits (red dashed curves). Inset illustrates measurement geometry; the arrow indicates changing with power LPB central frequency. (b) Reflectance spectra and fits at $k_x^{(2)}/k = 0.024$. (c) Extracted LPB spectral blueshifts for laser excitation at $k_x^{(1)}/k = 0.078$ (black squares) and $k_x^{(2)}/k = 0.024$ (red circles) as functions of incident fluence, together with corresponding linear fits (black and red lines).}
	\label{fig:Nonlinearity}
\end{figure*}

Mixing photonic modes with excitons in our hybrid MoSe$_2$/hBN/PCS structures leads to a dramatic enhancement of the associated optical nonlinearities.
We probe the underlying polariton--polariton interaction due to the excitonic contribution by measuring pump-dependent frequency shifts of polariton peaks in the reflectivity spectra.
Polariton modes are excited resonantly in both frequency and wavevector domains by $\sim 130$~fs laser pulses (Fig.~\ref{fig:Nonlinearity}a, inset), with incident fluence varying from 0.1~\textmu{}J/cm$^2$ to 3.0~\textmu{}J/cm$^2$ (see Methods).

Figs.~\ref{fig:Nonlinearity}a,b show the measured pump-dependent reflectivity spectra of the LPB resonance (solid black curves) for selected values of incident fluence, increasing from bottom to top, and for two different x-components of wavevector: $k_x^{(1)}/k = 0.078$ (a) and $k_x^{(2)}/k = 0.024$ (b). 
Due to exciton--exciton interaction, increasing with the density of created quasiparticles~\cite{Shahnazaryan2017}, lower-energy polariton resonance in the reflectivity spectra is continuously shifted with fluence towards higher energies.
We observe larger blueshifts (b) for wavevectors closer to the anticrossing condition $k_x^\mathrm{res} = 0.014$, as expected for the increasing exciton fraction in LPB associated with stronger polariton--polariton interaction. 

Fig.~\ref{fig:Nonlinearity}c shows the blueshift values for $k_x^{(1)}/k = 0.078$ (black squares) and $k_x^{(2)}/k = 0.024$ (red circles) extracted from Fano lineshape fitting (a-b, red dashed curves) for varying fluence, together with corresponding linear fits (black and red lines). 
Calculating polariton density $n_\mathrm{P}$ for each fluence (see Supplementary Note~5), we obtain the polariton--polariton interaction strength $g_\mathrm{P} = \mathrm{d}E_\mathrm{P}/\mathrm{d}n_\mathrm{P}$ of $g_\mathrm{P}(k_x^{(1)}) \sim 0.04$~\textmu{}eV$\cdot$\textmu{}m$^2$ and $g_\mathrm{P}(k_x^{(2)}) \sim 0.16$~\textmu{}eV$\cdot$\textmu{}m$^2$.
Further, from the $g_\mathrm{P}(k_x) \propto g_\mathrm{X}|X(k_x)|^4$ dependence on the Hopfield coefficient $X(k_x)$, which describes the exciton fraction in the polariton, we estimate the exciton--exciton interaction strength in our measurement as $g_\mathrm{X} \sim 1.0$~\textmu{}eV$\cdot$\textmu{}m$^2$.
This is of the same order as the theoretical estimate $g_\mathrm{X} \sim 1.6$~\textmu{}eV$\cdot$\textmu{}m$^2$, as well as the value $g_\mathrm{X} \sim 1.4$~\textmu{}eV$\cdot$\textmu{}m$^2$ we extract from a direct measurement of pump-dependent excitonic blueshifts in TM polarization (see Supplementary Note~6), and is considerably larger than the numbers recently reported for WS$_2$-based polaritons~\cite{Barachati2018}, where the estimation of exciton--exciton interaction strength was possibly uncertain due to local heating and higher-order interaction effects~\cite{Note1}.

As exciton densities in our experiment ($\leq 10^{12}$~cm$^{-2}$) are far below the Mott transition density ($\sim 10^{14}$~cm$^{-2}$), phase space filling effects are not important, and the observed polariton nonlinearity is dominated by exciton--exciton interaction.
Additionally, we note that the observed nonlinearities are fast at least on a 100~fs scale, providing future opportunities for developing polariton-based ultrafast modulators and switches.


In summary, we present the first experimental demonstration and investigation of optical BIC-based polaritonic excitations.
The formation of BIC-like polaritons in a hybrid system of a monolayer semiconductor interfaced with a photonic crystal slab, with suppressed radiation into far-field and line narrowing due to effective disorder averaging, extends the polariton--polariton interaction time for an enhancement of the nonlinear optical response.
In future, such ``dark" states can be accessed through near fields, using guided modes excited via grating-coupling, or by nonlinear frequency conversion.
With the strength of the underlying exciton--exciton interaction of $g_\mathrm{X} \sim 1.0$~\textmu{}eV$\cdot$\textmu{}m$^2$, our polaritons exhibit strong exciton-fraction-dependent optical nonlinearities that are fast on a 100~fs time scale.
Additionally, the planar geometry of our structure in principle allows straightforward fabrication of electrical contacts for electrostatic control of polaritons and associated interactions, while the use of atomically thin semiconductors allows room temperature operation.
Thus, the formation of BIC-based polaritons can enable not only significantly enhanced but also controllable and fast nonlinear optical responses in photonic crystal systems due to the strong excitonic interaction in monolayer semiconductors and open a new way to develop active and nonlinear all-optical on-chip devices. \\


\noindent
{\bf \large Methods}\\
\noindent
\textbf{Sample fabrication.}
Ta$_2$O$_5$ layers of 90~nm thickness were deposited on commercial SiO$_2$/Si substrates via e-beam assisted ion-beam sputtering.
Photonic crystal slabs were fabricated by patterning the Ta$_2$O$_5$ layers via combination of electron-beam lithography and by plasma etching to yield the following geometric parameters: pitch $p=500$~nm, groove width $w=220$~nm, and depth $d=90$~nm, characterized with scanning electron and atomic force microscopy measurements.
Large-area high quality flakes of multilayer hBN and monolayer MoSe$_2$ were mechanically exfoliated from commercial bulk crystals (HQ Graphene) and stacked vertically onto the photonic crystal sample surface via dry transfer to form a hybrid 1L MoSe$_2$/hBN/PCS structure.\\

\noindent
\textbf{Optical measurements.}
Angle-resolved reflectance spectroscopy was performed in a back focal plane setup with a slit spectrometer coupled to a liquid nitrogen cooled imaging CCD camera (Princeton Instruments SP2500+PyLoN), using white light from a halogen lamp for illumination (see Supplementary Fig.~S1).
For pump-dependent reflectivity measurements, the sample was excited by 130~fs pulses from a wavelength-tunable Ti:sapphire oscillator (Spectra-Physics, Tsunami, 80 MHz repetition rate) with wavevector control via laser beam positioning within the back focal plane of the objective.
A single-slit optical chopper with a duty cycle of 0.001 was used in the laser beam to avoid sample heating.
Angle-resolved photoluminescence (PL) measurements were performed in the same setup with off-resonant excitation by monochromatic light from a HeNe laser with wavelength $\lambda_\mathrm{exc} = 632.8$~nm.
The sample was mounted in an ultra low vibration closed cycle helium cryostat (Advanced Research Systems) and maintained at controllable temperature in a $7-300$~K range.
The cryostat was mounted on a precise xyz stage for sample positioning. 
Spatial filtering in the detection channel was used to selectively measure signals from the 1L MoSe$_2$/hBN/PCS sample area.\\

\noindent
\textbf{Data availability.}
The data that support the plots within this paper and other findings of this study are available from the corresponding author upon reasonable request.\\

\noindent
{\bf \large Acknowledgements}\\
\noindent
The authors acknowledge funding from Ministry of Education and Science of the Russian Federation Megagrant No. 14.Y26.31.0015, GosZadanie No. 3.1365.2017/4.6, and GosZadanie No. 3.8891.2017/8.9. A.I.T. and D.N.K. acknowledge UK EPSRC grant EP/P026850/1.
V.K. acknowledges financial support from ITMO Fellowship Program.
This work was carried out using equipment of the SPbU Resource Centers ``Nanophotonics'' and ``Nanotechnology''.
We thank M. Zhukov, A. Bukatin, and A. Chezhegov for assistance with sample characterization and A. Bogdanov for helpful discussion. \\


\noindent
{\bf \large Competing interests}\\
\noindent
The authors declare no competing financial interests.\\

\noindent
{\bf \large Additional information}\\
\noindent
\textbf{Supplementary Information} is available for this paper at $\ast\ast\ast$ 




\begin{thebibliography}{0}%
\makeatletter
\providecommand \@ifxundefined [1]{%
 \@ifx{#1\undefined}
}%
\providecommand \@ifnum [1]{%
 \ifnum #1\expandafter \@firstoftwo
 \else \expandafter \@secondoftwo
 \fi
}%
\providecommand \@ifx [1]{%
 \ifx #1\expandafter \@firstoftwo
 \else \expandafter \@secondoftwo
 \fi
}%
\providecommand \natexlab [1]{#1}%
\providecommand \enquote  [1]{``#1''}%
\providecommand \bibnamefont  [1]{#1}%
\providecommand \bibfnamefont [1]{#1}%
\providecommand \citenamefont [1]{#1}%
\providecommand \href@noop [0]{\@secondoftwo}%
\providecommand \href [0]{\begingroup \@sanitize@url \@href}%
\providecommand \@href[1]{\@@startlink{#1}\@@href}%
\providecommand \@@href[1]{\endgroup#1\@@endlink}%
\providecommand \@sanitize@url [0]{\catcode `\\12\catcode `\$12\catcode
  `\&12\catcode `\#12\catcode `\^12\catcode `\_12\catcode `\%12\relax}%
\providecommand \@@startlink[1]{}%
\providecommand \@@endlink[0]{}%
\providecommand \url  [0]{\begingroup\@sanitize@url \@url }%
\providecommand \@url [1]{\endgroup\@href {#1}{\urlprefix }}%
\providecommand \urlprefix  [0]{URL }%
\providecommand \Eprint [0]{\href }%
\providecommand \doibase [0]{http://dx.doi.org/}%
\providecommand \selectlanguage [0]{\@gobble}%
\providecommand \bibinfo  [0]{\@secondoftwo}%
\providecommand \bibfield  [0]{\@secondoftwo}%
\providecommand \translation [1]{[#1]}%
\providecommand \BibitemOpen [0]{}%
\providecommand \bibitemStop [0]{}%
\providecommand \bibitemNoStop [0]{.\EOS\space}%
\providecommand \EOS [0]{\spacefactor3000\relax}%
\providecommand \BibitemShut  [1]{\csname bibitem#1\endcsname}%
\let\auto@bib@innerbib\@empty
\end{thebibliography}%


\begin{thebibliography}{10}
\expandafter\ifx\csname url\endcsname\relax
  \def\url#1{\texttt{#1}}\fi
\expandafter\ifx\csname urlprefix\endcsname\relax\def\urlprefix{URL }\fi
\providecommand{\bibinfo}[2]{#2}
\providecommand{\eprint}[2][]{\url{#2}}

\bibitem{Marinica2008}
\bibinfo{author}{Marinica, D.}, \bibinfo{author}{Borisov, A.} \&
  \bibinfo{author}{Shabanov, S.}
\newblock \bibinfo{title}{Bound states in the continuum in photonics}.
\newblock \emph{\bibinfo{journal}{Phys. Rev. Lett.}}
  \textbf{\bibinfo{volume}{100}}, \bibinfo{pages}{183902}
  (\bibinfo{year}{2008}).

\bibitem{Hsu2016}
\bibinfo{author}{Hsu, C.~W.}, \bibinfo{author}{Zhen, B.},
  \bibinfo{author}{Stone, A.~D.}, \bibinfo{author}{Joannopoulos, J.~D.} \&
  \bibinfo{author}{Solja{\v{c}}i{\'c}, M.}
\newblock \bibinfo{title}{Bound states in the continuum}.
\newblock \emph{\bibinfo{journal}{Nat. Rev. Mater.}}
  \textbf{\bibinfo{volume}{1}}, \bibinfo{pages}{16048} (\bibinfo{year}{2016}).

\bibitem{Lee2012}
\bibinfo{author}{Lee, J.} \emph{et~al.}
\newblock \bibinfo{title}{Observation and differentiation of unique high-{Q}
  optical resonances near zero wave vector in macroscopic photonic crystal
  slabs}.
\newblock \emph{\bibinfo{journal}{Phys. Rev. Lett.}}
  \textbf{\bibinfo{volume}{109}}, \bibinfo{pages}{067401}
  (\bibinfo{year}{2012}).

\bibitem{Jin2018}
\bibinfo{author}{Jin, J.} \emph{et~al.}
\newblock \bibinfo{title}{Topologically enabled ultra-high-{Q} guided
  resonances robust to out-of-plane scattering}.
\newblock \emph{\bibinfo{journal}{arXiv preprint arXiv:1812.00892}}
  (\bibinfo{year}{2018}).

\bibitem{Krasikov2018}
\bibinfo{author}{Krasikov, S.}, \bibinfo{author}{Bogdanov, A.} \&
  \bibinfo{author}{Iorsh, I.}
\newblock \bibinfo{title}{Nonlinear bound states in the continuum of a
  one-dimensional photonic crystal slab}.
\newblock \emph{\bibinfo{journal}{Phys. Rev. B}} \textbf{\bibinfo{volume}{97}},
  \bibinfo{pages}{224309} (\bibinfo{year}{2018}).

\bibitem{Bulgakov2019}
\bibinfo{author}{Bulgakov, E.~N.} \& \bibinfo{author}{Maksimov, D.~N.}
\newblock \bibinfo{title}{Nonlinear response from optical bound states in the
  continuum}.
\newblock \emph{\bibinfo{journal}{arXiv preprint arXiv:1902.05782}}
  (\bibinfo{year}{2019}).

\bibitem{Walker2015}
\bibinfo{author}{Walker, P.} \emph{et~al.}
\newblock \bibinfo{title}{Ultra-low-power hybrid light--matter solitons}.
\newblock \emph{\bibinfo{journal}{Nat. Commun.}} \textbf{\bibinfo{volume}{6}},
  \bibinfo{pages}{8317} (\bibinfo{year}{2015}).

\bibitem{Sich2016}
\bibinfo{author}{Sich, M.}, \bibinfo{author}{Skryabin, D.~V.} \&
  \bibinfo{author}{Krizhanovskii, D.~N.}
\newblock \bibinfo{title}{Soliton physics with semiconductor
  exciton--polaritons in confined systems}.
\newblock \emph{\bibinfo{journal}{C. R. Phys.}} \textbf{\bibinfo{volume}{17}},
  \bibinfo{pages}{908--919} (\bibinfo{year}{2016}).

\bibitem{Kodigala2017}
\bibinfo{author}{Kodigala, A.} \emph{et~al.}
\newblock \bibinfo{title}{Lasing action from photonic bound states in
  continuum}.
\newblock \emph{\bibinfo{journal}{Nature}} \textbf{\bibinfo{volume}{541}},
  \bibinfo{pages}{196} (\bibinfo{year}{2017}).

\bibitem{Foley2014}
\bibinfo{author}{Foley, J.~M.}, \bibinfo{author}{Young, S.~M.} \&
  \bibinfo{author}{Phillips, J.~D.}
\newblock \bibinfo{title}{Symmetry-protected mode coupling near normal
  incidence for narrow-band transmission filtering in a dielectric grating}.
\newblock \emph{\bibinfo{journal}{Phys. Rev. B}} \textbf{\bibinfo{volume}{89}},
  \bibinfo{pages}{165111} (\bibinfo{year}{2014}).

\bibitem{Romano2018a}
\bibinfo{author}{Romano, S.} \emph{et~al.}
\newblock \bibinfo{title}{Surface-enhanced {R}aman and fluorescence
  spectroscopy with an all-dielectric metasurface}.
\newblock \emph{\bibinfo{journal}{J. Phys. Chem. C}}
  \textbf{\bibinfo{volume}{122}}, \bibinfo{pages}{19738--19745}
  (\bibinfo{year}{2018}).

\bibitem{Romano2018b}
\bibinfo{author}{Romano, S.} \emph{et~al.}
\newblock \bibinfo{title}{Optical biosensors based on photonic crystals
  supporting bound states in the continuum}.
\newblock \emph{\bibinfo{journal}{Materials}} \textbf{\bibinfo{volume}{11}},
  \bibinfo{pages}{526} (\bibinfo{year}{2018}).

\bibitem{Khitrova1999}
\bibinfo{author}{Khitrova, G.}, \bibinfo{author}{Gibbs, H.},
  \bibinfo{author}{Jahnke, F.}, \bibinfo{author}{Kira, M.} \&
  \bibinfo{author}{Koch, S.~W.}
\newblock \bibinfo{title}{Nonlinear optics of normal-mode-coupling
  semiconductor microcavities}.
\newblock \emph{\bibinfo{journal}{Rev. Mod. Phys.}}
  \textbf{\bibinfo{volume}{71}}, \bibinfo{pages}{1591} (\bibinfo{year}{1999}).

\bibitem{Dufferwiel2015}
\bibinfo{author}{Dufferwiel, S.} \emph{et~al.}
\newblock \bibinfo{title}{Exciton--polaritons in van der {W}aals
  heterostructures embedded in tunable microcavities}.
\newblock \emph{\bibinfo{journal}{Nat. Commun.}} \textbf{\bibinfo{volume}{6}},
  \bibinfo{pages}{8579} (\bibinfo{year}{2015}).

\bibitem{Liu2015}
\bibinfo{author}{Liu, X.} \emph{et~al.}
\newblock \bibinfo{title}{Strong light--matter coupling in two-dimensional
  atomic crystals}.
\newblock \emph{\bibinfo{journal}{Nat. Photonics}}
  \textbf{\bibinfo{volume}{9}}, \bibinfo{pages}{30} (\bibinfo{year}{2015}).

\bibitem{Lundt2016}
\bibinfo{author}{Lundt, N.} \emph{et~al.}
\newblock \bibinfo{title}{Room-temperature {T}amm-plasmon exciton-polaritons
  with a {WSe2} monolayer}.
\newblock \emph{\bibinfo{journal}{Nat. Commun.}} \textbf{\bibinfo{volume}{7}},
  \bibinfo{pages}{13328} (\bibinfo{year}{2016}).

\bibitem{Dibos2019}
\bibinfo{author}{Dibos, A.~M.} \emph{et~al.}
\newblock \bibinfo{title}{Electrically tunable exciton--plasmon coupling in a
  {WSe2} monolayer embedded in a plasmonic crystal cavity}.
\newblock \emph{\bibinfo{journal}{Nano Lett.}}  (\bibinfo{year}{2019}).

\bibitem{Zhang2018}
\bibinfo{author}{Zhang, L.}, \bibinfo{author}{Gogna, R.},
  \bibinfo{author}{Burg, W.}, \bibinfo{author}{Tutuc, E.} \&
  \bibinfo{author}{Deng, H.}
\newblock \bibinfo{title}{Photonic-crystal exciton-polaritons in monolayer
  semiconductors}.
\newblock \emph{\bibinfo{journal}{Nat. Commun.}} \textbf{\bibinfo{volume}{9}},
  \bibinfo{pages}{713} (\bibinfo{year}{2018}).

\bibitem{Gogna2019}
\bibinfo{author}{Gogna, R.}, \bibinfo{author}{Zhang, L.},
  \bibinfo{author}{Wang, Z.} \& \bibinfo{author}{Deng, H.}
\newblock \bibinfo{title}{Photonic crystals for controlling strong coupling in
  van der {W}aals materials}.
\newblock \emph{\bibinfo{journal}{Opt. Express}} \textbf{\bibinfo{volume}{27}},
  \bibinfo{pages}{22700--22707} (\bibinfo{year}{2019}).

\bibitem{Barachati2018}
\bibinfo{author}{Barachati, F.} \emph{et~al.}
\newblock \bibinfo{title}{Interacting polariton fluids in a monolayer of
  tungsten disulfide}.
\newblock \emph{\bibinfo{journal}{Nat. Nanotechnol.}}
  \textbf{\bibinfo{volume}{13}}, \bibinfo{pages}{906} (\bibinfo{year}{2018}).

\bibitem{Koshelev2018}
\bibinfo{author}{Koshelev, K.}, \bibinfo{author}{Sychev, S.},
  \bibinfo{author}{Sadrieva, Z.}, \bibinfo{author}{Bogdanov, A.} \&
  \bibinfo{author}{Iorsh, I.}
\newblock \bibinfo{title}{Strong coupling between excitons in transition metal
  dichalcogenides and optical bound states in the continuum}.
\newblock \emph{\bibinfo{journal}{Phys. Rev. B}} \textbf{\bibinfo{volume}{98}},
  \bibinfo{pages}{161113} (\bibinfo{year}{2018}).

\bibitem{Ferrier2011}
\bibinfo{author}{Ferrier, L.} \emph{et~al.}
\newblock \bibinfo{title}{Interactions in confined polariton condensates}.
\newblock \emph{\bibinfo{journal}{Phys. Rev. Lett.}}
  \textbf{\bibinfo{volume}{106}}, \bibinfo{pages}{126401}
  (\bibinfo{year}{2011}).

\bibitem{Sich2012}
\bibinfo{author}{Sich, M.} \emph{et~al.}
\newblock \bibinfo{title}{Observation of bright polariton solitons in a
  semiconductor microcavity}.
\newblock \emph{\bibinfo{journal}{Nat. Photon.}} \textbf{\bibinfo{volume}{6}},
  \bibinfo{pages}{50} (\bibinfo{year}{2012}).

\bibitem{Hsu2013}
\bibinfo{author}{Hsu, C.~W.} \emph{et~al.}
\newblock \bibinfo{title}{Observation of trapped light within the radiation
  continuum}.
\newblock \emph{\bibinfo{journal}{Nature}} \textbf{\bibinfo{volume}{499}},
  \bibinfo{pages}{188} (\bibinfo{year}{2013}).

\bibitem{Sadrieva2017}
\bibinfo{author}{Sadrieva, Z.~F.} \emph{et~al.}
\newblock \bibinfo{title}{Transition from optical bound states in the continuum
  to leaky resonances: role of substrate and roughness}.
\newblock \emph{\bibinfo{journal}{ACS Photonics}} \textbf{\bibinfo{volume}{4}},
  \bibinfo{pages}{723--727} (\bibinfo{year}{2017}).

\bibitem{Grepstad2013}
\bibinfo{author}{Grepstad, J.~O.} \emph{et~al.}
\newblock \bibinfo{title}{Finite-size limitations on quality factor of guided
  resonance modes in {2D} photonic crystals}.
\newblock \emph{\bibinfo{journal}{Opt. Express}} \textbf{\bibinfo{volume}{21}},
  \bibinfo{pages}{23640--23654} (\bibinfo{year}{2013}).

\bibitem{Hopfield1958}
\bibinfo{author}{Hopfield, J.}
\newblock \bibinfo{title}{Theory of the contribution of excitons to the complex
  dielectric constant of crystals}.
\newblock \emph{\bibinfo{journal}{Phys. Rev.}} \textbf{\bibinfo{volume}{112}},
  \bibinfo{pages}{1555} (\bibinfo{year}{1958}).

\bibitem{Whittaker1996}
\bibinfo{author}{Whittaker, D.} \emph{et~al.}
\newblock \bibinfo{title}{Motional narrowing in semiconductor microcavities}.
\newblock \emph{\bibinfo{journal}{Phys. Rev. Lett.}}
  \textbf{\bibinfo{volume}{77}}, \bibinfo{pages}{4792} (\bibinfo{year}{1996}).

\bibitem{Savona1997}
\bibinfo{author}{Savona, V.}, \bibinfo{author}{Piermarocchi, C.},
  \bibinfo{author}{Quattropani, A.}, \bibinfo{author}{Tassone, F.} \&
  \bibinfo{author}{Schwendimann, P.}
\newblock \bibinfo{title}{Microscopic theory of motional narrowing of
  microcavity polaritons in a disordered potential}.
\newblock \emph{\bibinfo{journal}{Phys. Rev. Lett.}}
  \textbf{\bibinfo{volume}{78}}, \bibinfo{pages}{4470} (\bibinfo{year}{1997}).

\bibitem{Kavokin1998}
\bibinfo{author}{Kavokin, A.~V.}
\newblock \bibinfo{title}{Motional narrowing of inhomogeneously broadened
  excitons in a semiconductor microcavity: {S}emiclassical treatment}.
\newblock \emph{\bibinfo{journal}{Phys. Rev. B}} \textbf{\bibinfo{volume}{57}},
  \bibinfo{pages}{3757} (\bibinfo{year}{1998}).

\bibitem{Whittaker1998}
\bibinfo{author}{Whittaker, D.}
\newblock \bibinfo{title}{What determines inhomogeneous linewidths in
  semiconductor microcavities?}
\newblock \emph{\bibinfo{journal}{Phys. Rev. Lett.}}
  \textbf{\bibinfo{volume}{80}}, \bibinfo{pages}{4791} (\bibinfo{year}{1998}).

\bibitem{Skolnick1998}
\bibinfo{author}{Skolnick, M.}, \bibinfo{author}{Fisher, T.} \&
  \bibinfo{author}{Whittaker, D.}
\newblock \bibinfo{title}{Strong coupling phenomena in quantum microcavity
  structures}.
\newblock \emph{\bibinfo{journal}{Semicond. Sci. Technol.}}
  \textbf{\bibinfo{volume}{13}}, \bibinfo{pages}{645} (\bibinfo{year}{1998}).

\bibitem{Rhodes2019}
\bibinfo{author}{Rhodes, D.}, \bibinfo{author}{Chae, S.~H.},
  \bibinfo{author}{Ribeiro-Palau, R.} \& \bibinfo{author}{Hone, J.}
\newblock \bibinfo{title}{Disorder in van der {W}aals heterostructures of {2D}
  materials}.
\newblock \emph{\bibinfo{journal}{Nat. Mater.}} \textbf{\bibinfo{volume}{18}},
  \bibinfo{pages}{541} (\bibinfo{year}{2019}).

\bibitem{Houdre1996}
\bibinfo{author}{Houdr{\'e}, R.}, \bibinfo{author}{Stanley, R.} \&
  \bibinfo{author}{Ilegems, M.}
\newblock \bibinfo{title}{Vacuum-field {R}abi splitting in the presence of
  inhomogeneous broadening: {R}esolution of a homogeneous linewidth in an
  inhomogeneously broadened system}.
\newblock \emph{\bibinfo{journal}{Phys. Rev. A}} \textbf{\bibinfo{volume}{53}},
  \bibinfo{pages}{2711} (\bibinfo{year}{1996}).

\bibitem{Walker2017}
\bibinfo{author}{Walker, P.} \emph{et~al.}
\newblock \bibinfo{title}{Dark solitons in high velocity waveguide polariton
  fluids}.
\newblock \emph{\bibinfo{journal}{Phys. Rev. Lett.}}
  \textbf{\bibinfo{volume}{119}}, \bibinfo{pages}{097403}
  (\bibinfo{year}{2017}).

\bibitem{Ajayi2017}
\bibinfo{author}{Ajayi, O.~A.} \emph{et~al.}
\newblock \bibinfo{title}{Approaching the intrinsic photoluminescence linewidth
  in transition metal dichalcogenide monolayers}.
\newblock \emph{\bibinfo{journal}{2D Mater.}} \textbf{\bibinfo{volume}{4}},
  \bibinfo{pages}{031011} (\bibinfo{year}{2017}).

\bibitem{Scuri2018}
\bibinfo{author}{Scuri, G.} \emph{et~al.}
\newblock \bibinfo{title}{Large excitonic reflectivity of monolayer {MoSe2}
  encapsulated in hexagonal boron nitride}.
\newblock \emph{\bibinfo{journal}{Phys. Rev. Lett.}}
  \textbf{\bibinfo{volume}{120}}, \bibinfo{pages}{037402}
  (\bibinfo{year}{2018}).

\bibitem{Martin2018}
\bibinfo{author}{Martin, E.~W.} \emph{et~al.}
\newblock \bibinfo{title}{Encapsulation narrows excitonic homogeneous linewidth
  of exfoliated {MoSe2} monolayer}.
\newblock \emph{\bibinfo{journal}{arXiv preprint arXiv:1810.09834}}
  (\bibinfo{year}{2018}).

\bibitem{Fang2019}
\bibinfo{author}{Fang, H.} \emph{et~al.}
\newblock \bibinfo{title}{Control of the exciton radiative lifetime in van der
  {W}aals heterostructures}.
\newblock \emph{\bibinfo{journal}{arXiv preprint arXiv:1902.00670}}
  (\bibinfo{year}{2019}).

\bibitem{Shahnazaryan2017}
\bibinfo{author}{Shahnazaryan, V.}, \bibinfo{author}{Iorsh, I.},
  \bibinfo{author}{Shelykh, I.} \& \bibinfo{author}{Kyriienko, O.}
\newblock \bibinfo{title}{Exciton-exciton interaction in transition-metal
  dichalcogenide monolayers}.
\newblock \emph{\bibinfo{journal}{Phys. Rev. B}} \textbf{\bibinfo{volume}{96}},
  \bibinfo{pages}{115409} (\bibinfo{year}{2017}).

\bibitem{Note1}
\bibinfo{note}{Experiments in Ref.~\cite {Barachati2018} are performed at 300
  K, so that heating effects might play a significant role at high powers,
  resulting in spectral redshifts reducing the observed blueshifts; in
  addition, due to the broad polariton lines, the energy shifts of $>2$~meV
  require quite high excitation densities, at which higher-order effects (three
  particle exciton--exciton interactions) become important and may lead to
  redshifts rather than blueshifts, reducing the overall observed nonlinear
  response}.

\end{thebibliography}

\begin{thebibliography}{10}
\expandafter\ifx\csname url\endcsname\relax
  \def\url#1{\texttt{#1}}\fi
\expandafter\ifx\csname urlprefix\endcsname\relax\def\urlprefix{URL }\fi
\providecommand{\bibinfo}[2]{#2}
\providecommand{\eprint}[2][]{\url{#2}}

\bibitem{Li1997}
\bibinfo{author}{Li, L.}
\newblock \bibinfo{title}{New formulation of the {F}ourier modal method for
  crossed surface-relief gratings}.
\newblock \emph{\bibinfo{journal}{J. Opt. Soc. Am. A}}
  \textbf{\bibinfo{volume}{14}}, \bibinfo{pages}{2758--2767}
  (\bibinfo{year}{1997}).

\bibitem{Houdre1996suppl}
\bibinfo{author}{Houdr{\'e}, R.}, \bibinfo{author}{Stanley, R.} \&
  \bibinfo{author}{Ilegems, M.}
\newblock \bibinfo{title}{Vacuum-field {R}abi splitting in the presence of
  inhomogeneous broadening: Resolution of a homogeneous linewidth in an
  inhomogeneously broadened system}.
\newblock \emph{\bibinfo{journal}{Phys. Rev. A}} \textbf{\bibinfo{volume}{53}},
  \bibinfo{pages}{2711} (\bibinfo{year}{1996}).

\bibitem{Whittaker1996suppl}
\bibinfo{author}{Whittaker, D.} \emph{et~al.}
\newblock \bibinfo{title}{Motional narrowing in semiconductor microcavities}.
\newblock \emph{\bibinfo{journal}{Phys. Rev. Lett.}}
  \textbf{\bibinfo{volume}{77}}, \bibinfo{pages}{4792} (\bibinfo{year}{1996}).

\bibitem{Mattuck1974}
\bibinfo{author}{Mattuck, R.~D.}
\newblock \emph{\bibinfo{title}{A guide to Feynman diagrams in the many-body
  problem}} (\bibinfo{publisher}{Courier Corporation}, \bibinfo{year}{1974}).

\bibitem{Ajayi2017suppl}
\bibinfo{author}{Ajayi, O.~A.} \emph{et~al.}
\newblock \bibinfo{title}{Approaching the intrinsic photoluminescence linewidth
  in transition metal dichalcogenide monolayers}.
\newblock \emph{\bibinfo{journal}{2D Mater.}} \textbf{\bibinfo{volume}{4}},
  \bibinfo{pages}{031011} (\bibinfo{year}{2017}).

\bibitem{Scuri2018suppl}
\bibinfo{author}{Scuri, G.} \emph{et~al.}
\newblock \bibinfo{title}{Large excitonic reflectivity of monolayer {M}o{S}e2
  encapsulated in hexagonal boron nitride}.
\newblock \emph{\bibinfo{journal}{Phys. Rev. Lett.}}
  \textbf{\bibinfo{volume}{120}}, \bibinfo{pages}{037402}
  (\bibinfo{year}{2018}).

\bibitem{Fang2019suppl}
\bibinfo{author}{Fang, H.} \emph{et~al.}
\newblock \bibinfo{title}{Control of the exciton radiative lifetime in van der
  {W}aals heterostructures}.
\newblock \emph{\bibinfo{journal}{arXiv preprint arXiv:1902.00670}}
  (\bibinfo{year}{2019}).

\bibitem{Barachati2018suppl}
\bibinfo{author}{Barachati, F.} \emph{et~al.}
\newblock \bibinfo{title}{Interacting polariton fluids in a monolayer of
  tungsten disulfide}.
\newblock \emph{\bibinfo{journal}{Nat. Nanotechnol.}}
  \textbf{\bibinfo{volume}{13}}, \bibinfo{pages}{906} (\bibinfo{year}{2018}).

\bibitem{Fan2003}
\bibinfo{author}{Fan, S.}, \bibinfo{author}{Suh, W.} \&
  \bibinfo{author}{Joannopoulos, J.~D.}
\newblock \bibinfo{title}{Temporal coupled-mode theory for the {F}ano resonance
  in optical resonators}.
\newblock \emph{\bibinfo{journal}{J. Opt. Soc. Am. A}}
  \textbf{\bibinfo{volume}{20}}, \bibinfo{pages}{569--572}
  (\bibinfo{year}{2003}).

\bibitem{Shahnazaryan2017suppl}
\bibinfo{author}{Shahnazaryan, V.}, \bibinfo{author}{Iorsh, I.},
  \bibinfo{author}{Shelykh, I.} \& \bibinfo{author}{Kyriienko, O.}
\newblock \bibinfo{title}{Exciton-exciton interaction in transition-metal
  dichalcogenide monolayers}.
\newblock \emph{\bibinfo{journal}{Phys. Rev. B}} \textbf{\bibinfo{volume}{96}},
  \bibinfo{pages}{115409} (\bibinfo{year}{2017}).

\bibitem{Li2015}
\bibinfo{author}{Li, J.}, \bibinfo{author}{Zhong, Y.} \&
  \bibinfo{author}{Zhang, D.}
\newblock \bibinfo{title}{Excitons in monolayer transition metal
  dichalcogenides}.
\newblock \emph{\bibinfo{journal}{J. Phys. Condens. Matter}}
  \textbf{\bibinfo{volume}{27}}, \bibinfo{pages}{315301}
  (\bibinfo{year}{2015}).

\end{thebibliography}

\pagebreak
\widetext
\begin{center}
\textbf{\large Supplementary Information: Nonlinear polaritons in monolayer semiconductor coupled to optical bound states in the continuum}
\end{center}
\setcounter{equation}{0}
\setcounter{figure}{0}
\setcounter{table}{0}
\setcounter{page}{1}
\makeatletter
\renewcommand{\thepage}{S\arabic{page}}
\renewcommand{\theequation}{S\arabic{equation}}
\renewcommand{\thefigure}{S\arabic{figure}}
\renewcommand{\bibnumfmt}[1]{[S#1]}
\renewcommand{\citenumfont}[1]{S#1}

\begin{figure}[b!]
	\includegraphics[width=0.9\columnwidth]{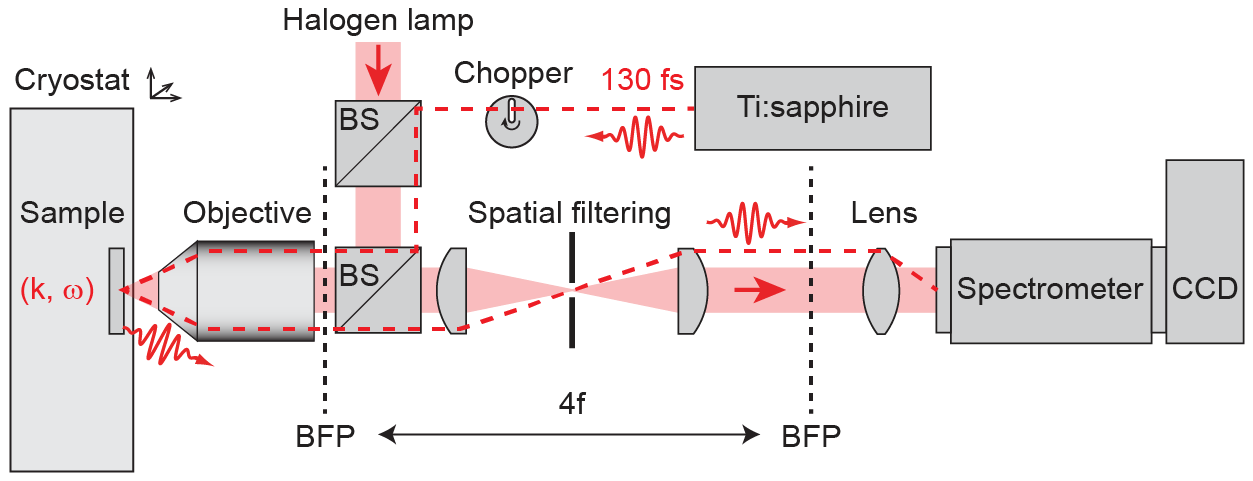}
	\caption{Schematic of experimental setup for angle-resolved reflectivity measurements. A halogen lamp is used for measurements in the linear optical regime, and a tunable Ti:sapphire oscillator with 130~fs pulses for nonlinear experiments. A single-slit optical chopper with 0.001 duty cycle is used to reduce sample heating. Angle-resolved data are collected from the back focal plane (BFP) of a microscope objective with subsequent filtering in real space (in a 4f configuration) to eliminate unwanted background signal. Samples are mounted in a closed-cycle He-free cryostat with micrometric positioning along 3 spatial axes. Resonant laser excitation is achieved by focusing laser pulses in BFP with a corresponding $\sim 200$~\micron~focal spot and tuning Ti:sapphire oscillator into resonance with a polariton mode. All signals are detected with a slit spectrometer ($f=500$~mm, 600~g/mm) and liquid nitrogen cooled CCD camera. BS: beamsplitter.}
	\label{fig:Setup}
\end{figure}

\begin{figure}[tb]
	\includegraphics[width=0.65\columnwidth]{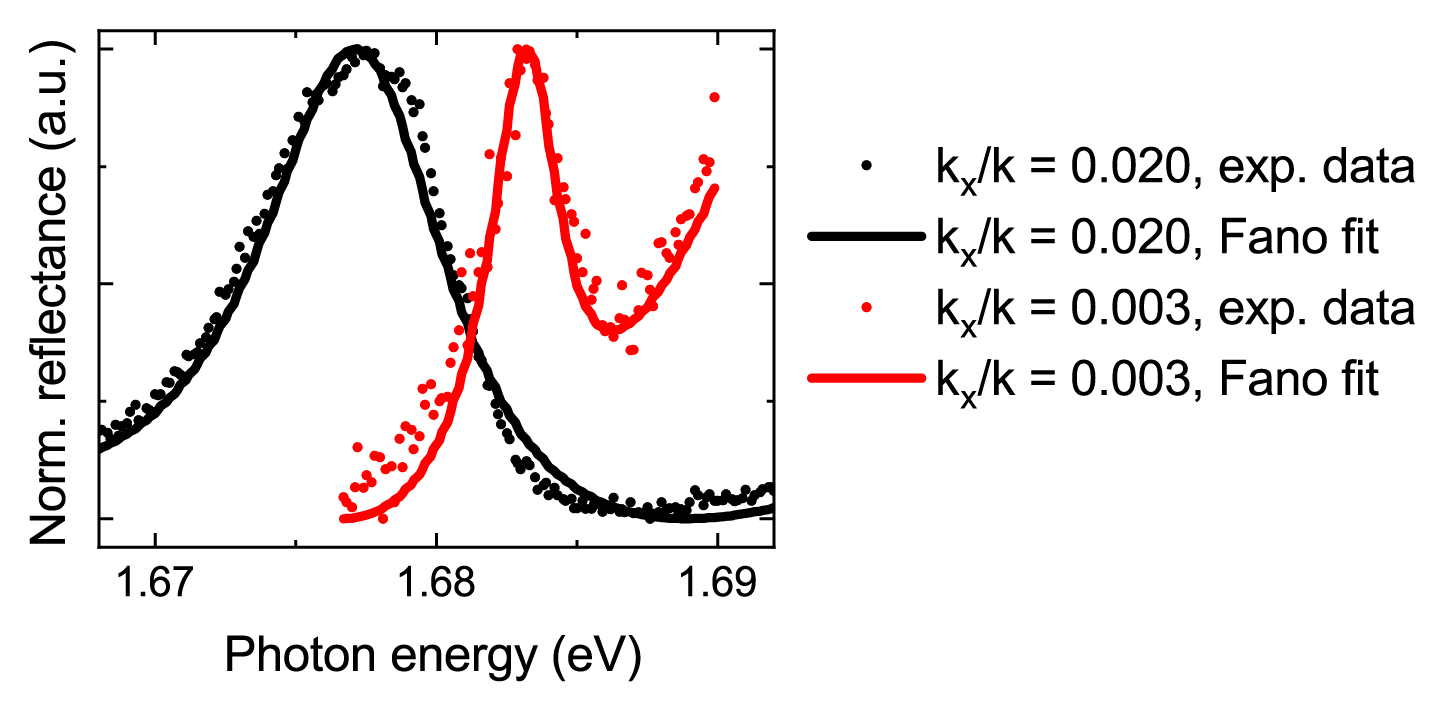}
	\caption{Normalized differential reflectance spectra from a PCS sample for two selected wavevectors: $k_x/k = 0.020$ (black) and $k_x/k = 0.003$ (red). Experimental data are indicated with dots, and lines represent Fano-like fits. For different wavevectors, Fano parameter varies, resulting in different peak shapes. The shoulder on the blue side is due to the higher energy symmetric mode of the photonic crystal.}
	\label{fig:Fano}
\end{figure}

\begin{figure}[tb]
	\includegraphics[width=\columnwidth]{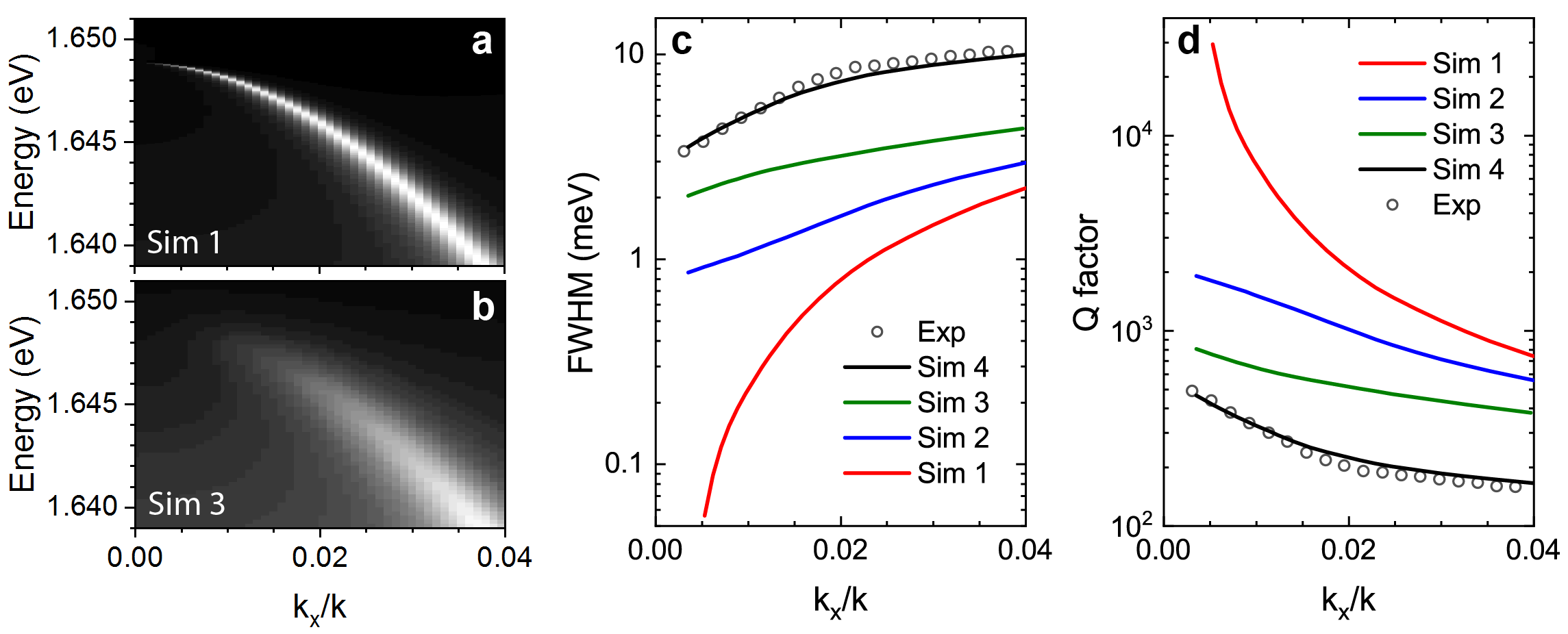}
    \caption{Influence of losses on the $m=2$ lower-energy antisymmetric mode in PCS: simulated reflectivity as a function of photon energy and x-component of wavevector without losses in Ta$_2$O$_5$ (a) and corresponding simulation result when taking into account intrinsic losses in Ta$_2$O$_5$ together with additional scattering losses $\delta n = 0.002i$. Linewidth (c) and quality factor (d) of the $m=2$ lower-energy antisymmetric mode obtained in experiment (Exp, circles) and in simulations (Sim 1-4, lines). Sim 1 (red): no losses (except leakage into Si), Sim 2 (blue): only intrinsic losses in Ta$_2$O$_5$ are considered, Sim 3 (green): scattering-related losses are added, Sim 4 (black): intrinsic losses, scattering-related losses, and line broadening due to the finite sample size are taken into account.}
	\label{fig:Qfactor}
\end{figure}

\begin{figure}[tb]
	\includegraphics[width=\columnwidth]{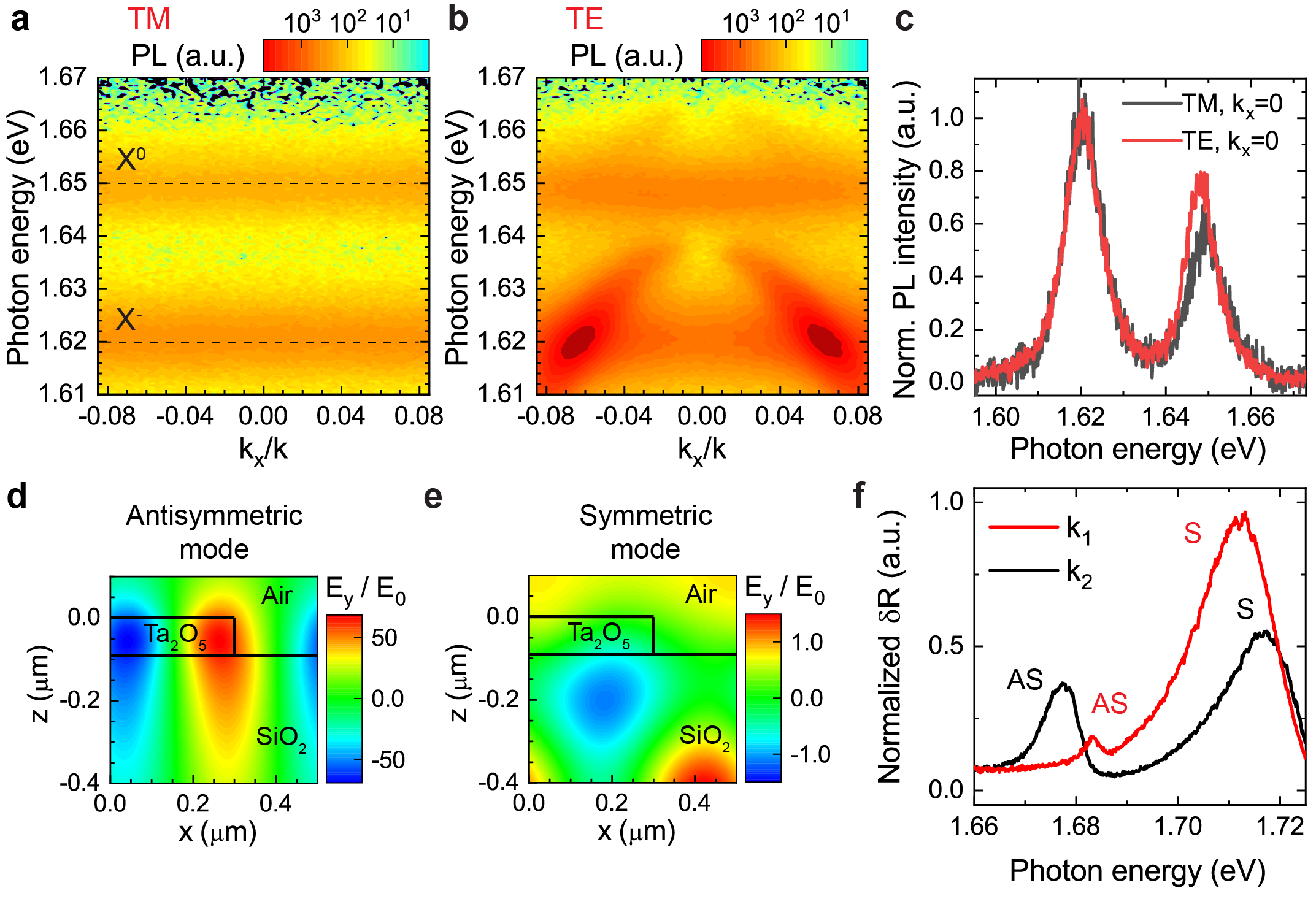}
	\caption{Angle resolved PL spectra detected in TM (a) and TE (b) polarization. TM spectra show angle-independent neutral exciton (X$^0$) and trion (X$^-$) emission. TE spectra (b) in addition show upper and lower polariton branches. PL spectra at zero angle (c) normalized on the trion peak intensity show slightly enhanced and redshifted neutral exciton emission in TE polarization, due to weak coupling to the symmetric mode of the PCS. Symmetric (S) and antisymmetric (AS) modes in PCS: spatial distribution of the y-component of electric field for S (a) and AS (b). Normalized differential reflectance spectra for two different wavevectors (k$_1 = 0.003$k, red, and k$_2 = 0.02$k, black), with distinct S and AS peaks (c).}
	\label{fig:UncoupledX}
\end{figure}

\section*{Supplementary Note 1. BIC Q-factor simulations}
With diverging to infinity radiative Q-factors of optical BICs in photonic crystal slabs (PCSs), their total Q-factors are limited by nonradiative terms, including material absorption, surface roughness induced scattering, leakage through the oxide layer into the Si substrate, and losses due to the finite sample size.
To estimate Q-factors achievable in our experimental system, we perform simulations of their optical response using the Fourier modal method (FMM)~\cite{Li1997}.
We first simulate a patterned Ta$_2$O$_5$/SiO$_2$/Si structure with the experimental parameters (Ta$_2$O$_5$ thickness of 90~nm, SiO$_2$ thickness of 1~\micron, pitch $p=500$~nm, groove width $w=220$~nm, and depth $d=90$~nm), but ignoring losses in Ta$_2$O$_5$ (setting the imaginary part of its refractive index to zero).

The resulting dispersion of the lower-energy $m=2$ antisymmetric mode is shown as angle-resolved reflectivity spectra in Fig.~\ref{fig:Qfactor}a and labeled as Sim~1.
Fitting the reflectivity spectra for different wavevectors with Fano-like shape as discussed in the main text, we obtain the wavevector dependence of the mode linewidth plotted in (c) with a red curve.
The corresponding Q-factor of the mode is plotted in (d) as a red curve and reveals a sharp increase towards the $\Gamma$ point, with Q-factors reaching $Q \sim 3\times 10^4$ at $k_x/k = 0.005$ and growing further towards $Q \sim 10^5$.

In contrast, when taking into account the absorption losses in Ta$_2$O$_5$, the Q-factors are limited by $Q \sim 2\times 10^3$, as indicated in (c) and (d) with a blue curve (labeled as Sim~2).
To describe our experimental data, we account for additional scattering losses due to surface roughness via an extra imaginary term in the refractive index of Ta$_2$O$_5$, $\delta n \sim 0.002i$ (Sim~3), with linewidth and Q-factor shown in (c) and (d) with green curves, and dispersion shown in (b).
Then, a convolution of the resulting lineshape with a Gaussian function that accounts for the finite sample size of $\sim 100$~\micron~yields linewidth and corresponding Q-factor as shown with a black curve in (c) and (d).
Experimental values (open black circles) are also plotted for reference.
Overall, the absorption in Ta$_2$O$_5$ is seen to be the major limiting factor for the BIC linewidth in this case.


\section*{Supplementary Note 2. Measurements in TE vs. TM polarization}
In TM polarization (E-field perpendicular to the grooves), the MoSe$_2$ exciton is far detuned from the photonic crystal mode.
As shown in Fig.~\ref{fig:UncoupledX}a, PL spectra detected in TM polarization reveal only wavevector-independent uncoupled neutral exciton (X$_0$) and trion (X$^-$) emission.
In TE polarization, in addition to the upper and lower polariton branches as well as the trion peak, our experimental PL spectra (b) show emission close to the frequency of the uncoupled neutral exciton in 1L MoSe$_2$ with a slight redshift of $\sim 1$~meV and enhancement as compared to the spectra measured in TM polarization (c).

This behavior is due to coupling to the $m=3$ symmetric PCS mode lying at higher energies.
While the strongly coupled to the exciton antisymmetric mode exhibits a tightly confined and enhanced electric field distribution (d), the symmetric mode is less confined, resulting in a weaker coupling, which, together with a larger detuning from the exciton resonance, leads to only a small ``Lamb'' redshift and slight enhancement of the exciton emission.
The corresponding narrow antisymmetric (AS) and broader symmetric (S) peaks observed in the PCS reflectivity spectra are shown in (f) for two different wavevectors.

\section*{Supplementary Note 3. Polariton linewidth model}
Since the Rabi splitting well exceeds the \textit{inhomogeneous} exciton linewidth in our case, the polariton linewidth close to the anticrossing point ($\omega_\mathrm{C}(k_x^0) = \omega_\mathrm{X}$) will be defined through only the \textit{homogeneous} contribution $\gamma_\mathrm{X}^\mathrm{h}$ to the exciton linewidth~\cite{Houdre1996suppl}.
We describe the experimentally obtained polariton linewidth with a phenomenological model taking into account both homogeneous and inhomogeneous exciton linewidth, $\gamma_\mathrm{X}^\mathrm{h}$ and $\gamma_\mathrm{X}^\mathrm{inh}$.
The effect of motional narrowing~\cite{Whittaker1996suppl} results in the dependence of the polariton linewidth on the effective mass~\cite{Mattuck1974}, which is incorporated in our model as
\begin{align}
    \gamma_{\pm}=|C_{\pm}|^2\gamma_\mathrm{C}+|X_{\pm}|^2\gamma_\mathrm{X}^\mathrm{h}+|X_{\pm}|^2\frac{M_{\pm}}{m_\mathrm{x}}\gamma_\mathrm{X}^\mathrm{inh},
\end{align}
where $M_{\pm}$, $m_\mathrm{X}$ are effective masses of the polaritons and exciton, respectively, and  $C_{\pm}$, $X_{\pm}$ are the Hopfield coefficients:
\begin{align}
|C_{\pm}|^2=\frac{\hbar^2\Omega_R^2}{|\hbar^2\Omega_R^2+(\omega_{\pm}-\omega_\mathrm{C})^2|},\quad |X_{\pm}|^2=1-|C_{\pm}|^2.
\end{align}
An agreement between the model and experimental LPB linewidth is reached for $\gamma_\mathrm{X}^\mathrm{h} \sim 1.0$~meV, which is close to recently reported values~\cite{Ajayi2017suppl, Scuri2018suppl, Fang2019suppl} for the low-temperature radiative decay rate of excitons in monolayer MoSe$_2$ of $\sim 1-2$~meV, implying that the homogeneous non-radiative contribution to MoSe$_2$ exciton linewidth at 7~K is small.
Assuming a Voigt profile of the exciton line at 7~K, we estimate the inhomogeneous linewidth of $\gamma_\mathrm{X}^\mathrm{inh} \sim 8.4$~meV.
The total homogeneous linewidth $\gamma_\mathrm{X}^\mathrm{h} = \gamma_\mathrm{X}^\mathrm{nrad} + \gamma_\mathrm{X}^\mathrm{rad}$ is then used to fit our experimentally measured dispersion with the coupled oscillator model at each temperature.

\begin{figure}[tb]
	\includegraphics[width=\columnwidth]{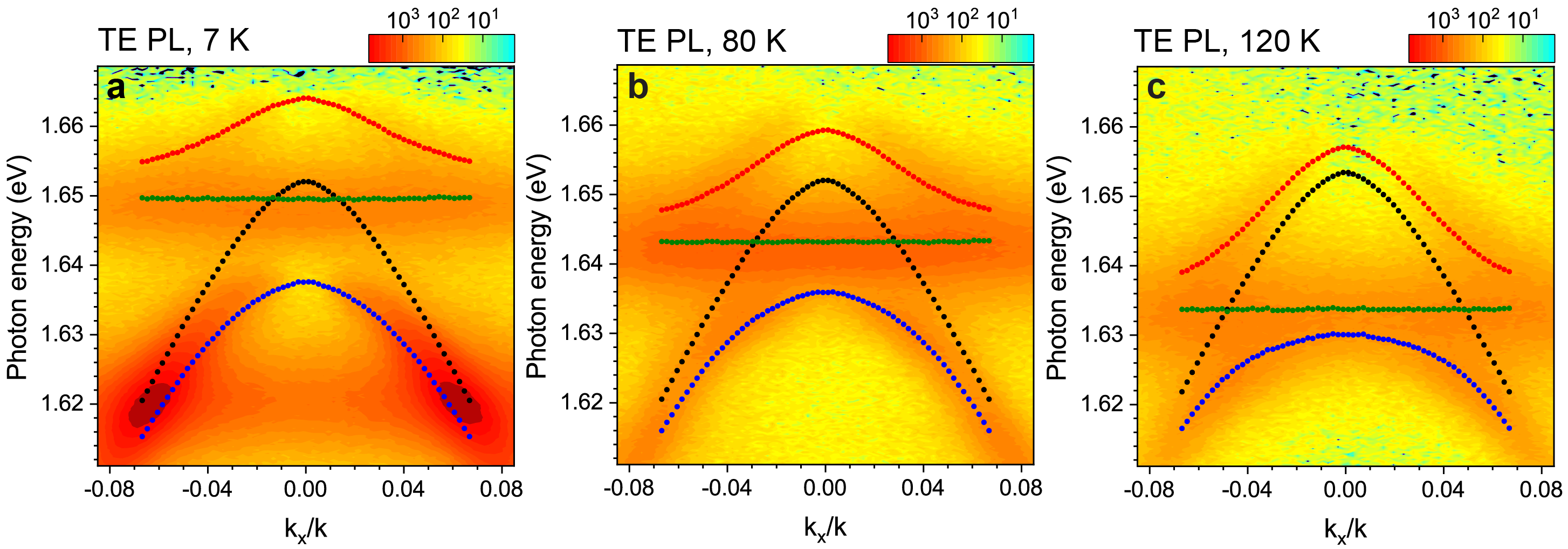}
	\caption{Temperature tuning of the polariton dispersion, as observed from selected PL spectra in TE polarization at 7~K (a), 80~K (b), and 120~K (c). Black: PCS mode; green: neutral exciton; red: upper polariton branch; blue: lower polariton branch. Spectral peak positions are extracted at each angle by fitting PL spectra with Lorentzian functions.}
	\label{fig:TemperaturePL}
\end{figure}

\begin{figure}[tb]
	\includegraphics[width=\columnwidth]{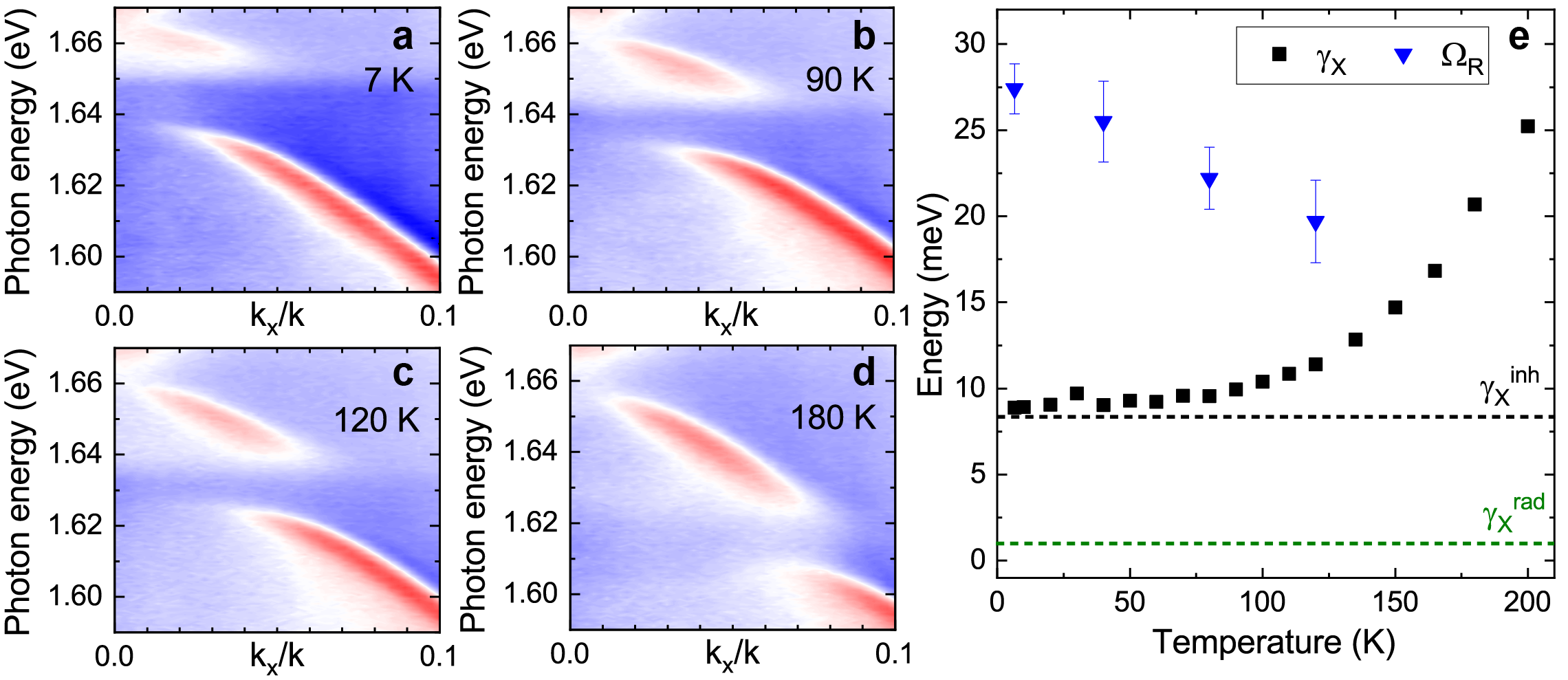}
	\caption{Temperature dependence of MoSe$_2$/hBN/PCS optical response: (a-c) Selected angle-resolved reflectance spectra measured at 7~K, 90~K, 120~K, and 180~K. (e) Extracted Rabi splitting (blue triangles) and total exciton linewidth (black squares) for temperature in a range of $7-200$~K. Estimated radiative exciton linewidth $\gamma_\mathrm{X}^\mathrm{rad}$ and inhomogeneous linewidth due to disorder $\gamma_\mathrm{X}^\mathrm{inh}$ are indicated with green and black dashed lines, respectively.}
	\label{fig:Tdependence}
\end{figure}

\section*{Supplementary Note 4. Variable temperature measurements}
With increasing temperature, the neutral MoSe$_2$ exciton shifts towards lower energies, effectively tuning the polariton dispersion.
Fig.~\ref{fig:TemperaturePL} shows experimental angle-resolved PL spectra for 3 selected temperatures of 7~K (a), 80~K (b), and 120~K (c), with the uncoupled PCS mode dispersion (black), uncoupled exciton (green), and upper and lower polariton branches (red and blue, respectively), as obtained by fitting the PL data with the coupled oscillator model.
The temperature dependence of the polariton dispersion is also observed in reflectivity spectra (Fig.~\ref{fig:Tdependence}a-d).
We extract a temperature-dependent coupling constant $g(T)$ (e, blue triangles) and compare it with the exciton linewidth (black squares).
The clear signatures of strong coupling persist up to at least 150~K where exciton broadening starts to blur the observed anticrossing behavior.

\section*{Supplementary Note 5. Polariton nonlinearity}
The nonlinear polaritonic shift $\Delta E_{pol}$ can be estimated as~\cite{Barachati2018suppl}
\begin{align}
    \Delta E_{pol}\approx g_\mathrm{X}|X|^4\int d^2\mathbf{r}|\psi(\mathbf{r})|^4\left[\int d^2\mathbf{r}|\psi(\mathbf{r})|^2 \right]^{-1},
\end{align}
where $g_\mathrm{X}$ is the exciton--exciton interaction constant, $X$ is the exciton Hopfield coefficient for the corresponding polariton mode, and $\psi(r)$ is the polariton wavefunction.
For the pulsed excitation case, $\psi$ is time-dependent; for the nonlinear shift modelling, the wavefunction is evaluated at the moment of time when it acquires the largest value.

For a pulsed pump with spectral width $\gamma_{pump}$ and polaritonic mode with radiative decay rate $\gamma_0$, the polaritonic wavefunction can be estimated (in the rectangular pulse shape approximation) as
\begin{align}
    \psi(\mathbf(r))\approx\psi_{sat}(\mathbf{r})(1-\exp[-\gamma_0/\gamma_{pump}]),
\end{align}
where $\psi_{sat}$ is the saturated wavefunction, which corresponds to the case of continuous wave excitation.

The saturated wavefunction can be estimated as
\begin{align}
    \psi_{sat}(x)=\sqrt{\frac{W_{max}}{c \hbar\omega_{p}}}\phi(x)=\sqrt{\frac{W_{max} }{c \hbar\omega_{p}}}\sqrt{\int \mathcal{E}^2(x,z)dz},
\end{align}
where $W_{max}$ is the peak power density, $\omega_p$ is the polariton frequency, and $\mathcal{E}$ is the dimensionless field distribution of the eigenmode when it is excited by a plane wave of the unity amplitude and frequency $\omega_p$. 
$\mathcal{E}$ is calculated numerically using the Fourier modal method (FMM).
It should be noted though that the general result of the temporal coupled mode theory~\cite{Fan2003} states that $\mathcal{E}$ is proportional to ${\gamma_0}^{-1/2}$.

The average polariton density is given by
\begin{align}
    \langle n_{pol}\rangle=\langle|\psi(x)|^2\rangle=\frac{W_{max} (1-\exp[-\gamma_0/\gamma_{pump}])^2}{c \hbar\omega_{p}}\frac{\int_0^D dx |\phi(x)|^4}{\int_0^D dx |\phi(x)|^2},
\end{align}
where $D$ is the period of the structure.
The upper bound for the polariton density for the maximum fluence of $3~\mu \mathrm{J~cm}^{-2}$ can be estimated as $n_{pol}^{max} \sim 10^{12}~\mathrm{cm}^{-2}$.
Thus, we are quite far from the Mott transition densities $a_b^{-2}\approx 10^{14}~\mathrm{cm}^{-2}$, and phase space filling effects can be neglected.

Finally, the polariton shift is given by
\begin{align}
    \Delta E_{pol}=g_\mathrm{X}|X|^4 n_{pol}=g_\mathrm{X}|X|^4\frac{W_{max} (1-\exp[-\gamma_0/\gamma_{pump}])^2}{c \hbar\omega_{p}}\frac{\int_0^D dx |\phi(x)|^4}{\int_0^D dx |\phi(x)|^2}.
\end{align}
The values of $|X|^2$ and $\gamma_0/\gamma_{pump}$ can be calculated for each wavevector.
For example, for the two wavevectors corresponding to the data in Fig.~4 in the main text, these values are estimated as $|X|^2=0.40, \gamma_0/\gamma_{pump}\approx 1$ for $k/k_0=0.024$ and $|X|^2=0.12, \gamma_0/\gamma_{pump}\approx 2$ for $k/k_0=0.078$.
At the maximum fluence $3~\mu J/\mathrm{cm}^{-2}$ this corresponds to polariton densities $3.1\times 10^4$ $\mu$m$^{-2}$ for $k/k_0=0.024$ and $4.5\times 10^4$ $\mu$m$^{-2}$ for $k/k_0=0.078$.
From the dependence of the polariton shift on the exciton Hopfield coefficient $X$, the exciton--exciton interaction strength can then be evaluated as $g_\mathrm{X} \sim 1.0~\mu\mathrm{eV}\cdot\mu\mathrm{m}^2$.

\section*{Supplementary Note 6. Exciton nonlinearity}

\begin{figure}[tb]
	\includegraphics[width=0.65\columnwidth]{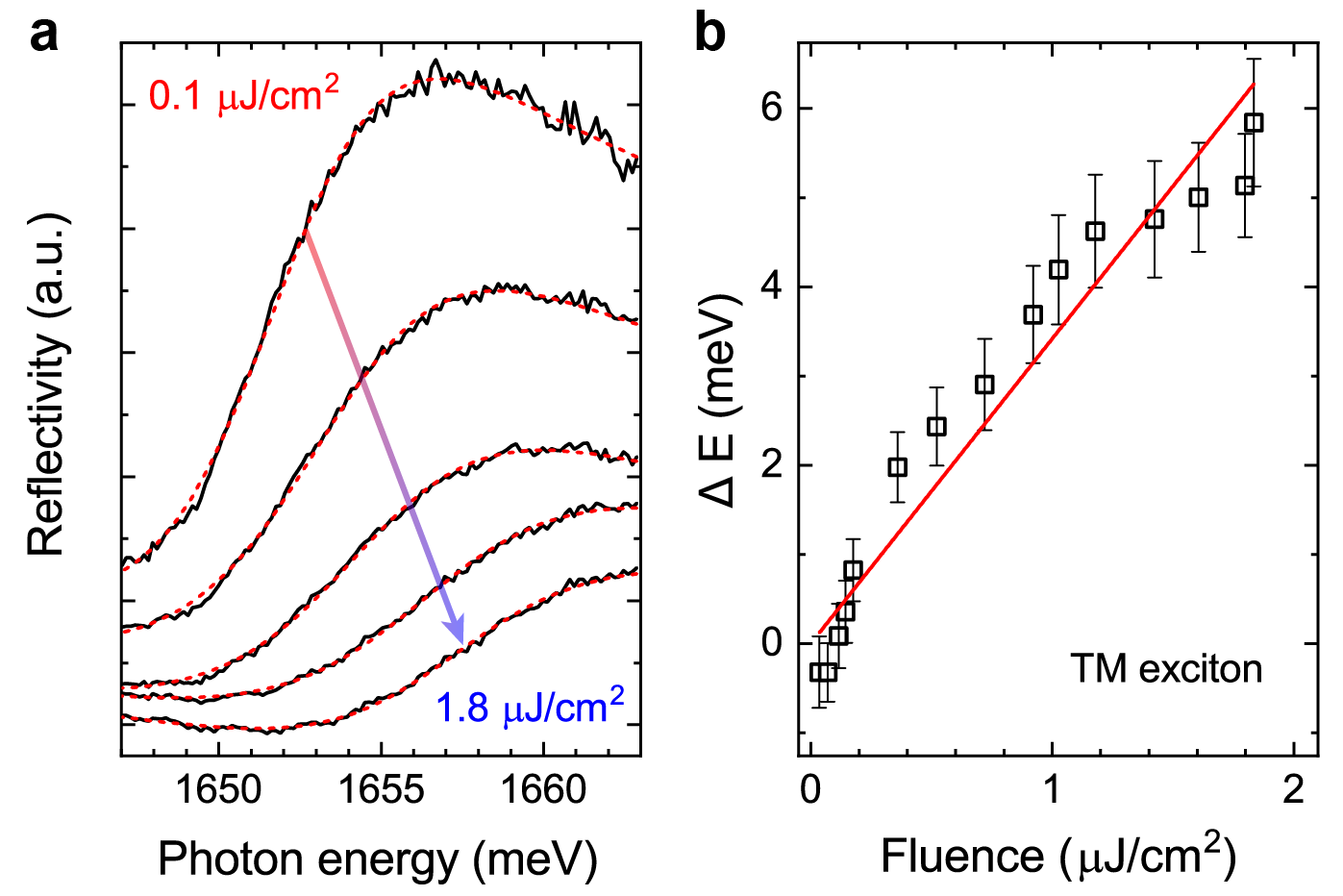}
	\caption{Nonlinearity of excitons in TM polarization. (a) Measured exciton reflectance spectra (solid black curves) under resonant illumination with laser pulses for varying fluence, together with corresponding Fano fits (red dashed curves). The arrow indicates changing with power exciton resonance frequency. (b) Extracted excitonic spectral blueshifts as a function of incident laser fluence, together with corresponding linear fit (black line).}
	\label{fig:NonlinearityX}
\end{figure}

We measure the pump-dependent blueshift of the exciton resonance in reflectance with TM-polarized excitation.
The corresponding reflectance spectra are shown in Fig.~\ref{fig:NonlinearityX}a for selected values of fluence varying from 0.1~\textmu{}J/cm$^2$ (top) to 1.8~\textmu{}J/cm$^2$ (bottom).
We fit the experimentally measured spectra with a Fano-like shape to obtain pump-dependent blueshift shown in Fig.~\ref{fig:NonlinearityX}b with open squares.
The exciton concentration for each fluence can be evaluated in the linear regime (omitting the effects of phase space filling) from the differential equation for the exciton annihilation operator $b$:
\begin{align}
    \frac{db}{dt}=-i\omega_X b-\frac{1}{2}(\gamma_0+\gamma_X^h)b +\frac{\sqrt{\gamma_0}}{2}(1+r)a_0\exp[-\frac{t^2}{2\tau_p^2}]\exp[-i\omega t],
\end{align}
where $\tau_p$ is the pulse duration, $r$ is the reflection coefficient of the substrate, and $a_0$ is the square root of the  number of photons passing through the monolayer per unit time per unit area, which is related to the peak incident power density $W$ as $a_0=\sqrt{W/(\hbar\omega)}$.
We have substituted the bosonic operator with the complex amplitude.
This differential equation can be solved explicitly yielding for $N=|b|^2$
\begin{align}
   & N=\nonumber \\&=|1+r|^2\frac{|a_0|^2}{\gamma_0}(\gamma_0\tau_p)^2\frac{\pi}{4}\left|\mathrm{erf}\left[\frac{t}{\sqrt{2}\tau_p}+\frac{i\tau_p}{\sqrt{2}}(\delta+i\gamma/2)\right]+1\right|^2\exp\left[-\tau_p^2(\delta^2-\gamma^2/4)-\gamma t\right],
\end{align}
where $\delta=\omega-\omega_X$, $\gamma=\gamma_0+\gamma_X^h$, and $t=0$ is at the pulse centre.
To account for the inhomogeneous broadening, we should convolve the expression for N with the gaussian distribution of the exciton frequency with the width $\gamma_{X}^{\mathrm{inh}}$:
\begin{align}
\tilde{N}=\frac{1}{\sqrt{\pi}\gamma_{X}^{\mathrm{inh}}}\int d\delta N(\delta)e^{-\delta^2/\gamma_{X}^{\mathrm{inh}}}
\end{align}    
This integral can not be taken analytically.
For $\gamma_0\approx\gamma_X^h=1$~meV, $\tau_p=130$~fs, and $\gamma_{X}^{\mathrm{inh}}=8.4$~meV, the numerical integration returns a function of time with a maximum at $\xi_{max}\approx 0.8$.

In the case of TM polarization the reflection coefficient is almost constant across the pulse bandwidth, and $r\approx -0.25-0.35i$, yielding $|1+r|^2=0.68$.
$a_0$ can be expressed through fluence $F$ as $|a_0|= \sqrt{F/\tau_p/\sqrt{\pi}\hbar\omega_X}$.

Finally, the exciton concentration N for fluence $F$ will be
\begin{align}
    \tilde{N}[\mu \mathrm{m} ^{-2}]\approx 2800 F[\mu \mathrm{J\cdot cm}^{-2}].
\end{align}
Then $g_\mathrm{X}$ can be evaluated as $g_\mathrm{X}=1.4~ \mu \mathrm{eV}\cdot\mu \mathrm{m}^2$.

We compare the obtained $g_\mathrm{X}$ value with a theoretical estimate~\cite{Shahnazaryan2017suppl} of $g_\mathrm{X} \sim 2 E_{1s}r_{1s}^2 \sim 1.6~\mu \mathrm{eV}\cdot\mu \mathrm{m}^2$, where $E_{1s} \sim 0.44$~eV is the $1s$ exciton binding energy in monolayer MoSe$_2$ and $r_{1s} \sim 1.36$~nm is the corresponding exciton Bohr radius~\cite{Li2015}.



\end{document}